\documentclass{aastex63}

\usepackage{amsmath}
\usepackage{natbib}

\usepackage{anyfontsize}

\usepackage{indentfirst}
\usepackage{comment}
\usepackage{multirow} 
\usepackage{booktabs}
\usepackage{bigints}
\usepackage{mathrsfs,amsmath} 
\usepackage{makecell}

\usepackage[normalem]{ulem}
\useunder{\uline}{\ul}{}
\usepackage{hyperref}
\usepackage{lineno}

\hypersetup{linkcolor=cyan,citecolor=magenta,filecolor=yellow,urlcolor=blue}

\hypersetup{linkcolor=magenta, citecolor=cyan, filecolor=yellow, urlcolor=blue}

\received{ddmmyyyy}
\revised{\today}
\accepted{ddmmyyyy}
\published{ddmmyyyy}

\begin{document}

\title{Revisiting the Spectral-Energy Correlations of GRBs with {\it Fermi} Data I:\\ Model-wise Properties}

\correspondingauthor{Liang Li}
\email{liang.li@icranet.org}


\author[0000-0002-1343-3089]{Liang Li}
\affiliation{ICRANet, Piazza della Repubblica 10, I-65122 Pescara, Italy}
\affiliation{INAF -- Osservatorio Astronomico d'Abruzzo, Via M. Maggini snc, I-64100, Teramo, Italy}
\affiliation{Dip. di Fisica and ICRA, Sapienza Universita di Roma, Piazzale Aldo Moro 5, I-00185 Rome, Italy}

\begin{abstract}

Gamma-ray bursts (GRBs) exhibit a diversity of spectra. Several spectral models (e.g., Band, cutoff power-law, and blackbody) and their hybrid versions (e.g., Band+blackbody) have been widely used to fit the observed GRB spectra. Here, we attempt to collect all the bursts detected by {\it Fermi}-GBM with known redshifts from July 2008 to May 2022, motivated to (i) provide a parameter catalog independent from the official \emph{Fermi}/GBM team and (ii) achieve a ``clean" model-based GRB spectral-energy correlation analysis. A nearly complete GRB sample was created, containing 153 such bursts (136 long gamma-ray bursts and 17 short gamma-ray bursts). Using the sample and by performing detailed spectral analysis and model comparisons, we investigate two GRB spectral-energy correlations: the cosmological rest-frame peak energy ($E_{\rm p,z}$) of the $\nu F_\nu$ prompt emission spectrum correlated with (i) the isotropic-bolometric-equivalent emission energy $E_{\gamma, \rm iso}$ (the Amati relation), and (ii) the isotropic-bolometric-equivalent peak luminosity $L_{\rm p, iso}$ (the Yonetoku relation). From a linear regression analysis, a tight correlation between $E_{\rm p,z}$ and $E_{\gamma, \rm iso}$ (and $L_{\gamma,\rm iso}$) is found for both the Band-like and CPL-like bursts. More interestingly, the CPL-like bursts do not fall on the Band-like burst Amati and Yonetoku correlations, suggesting distinct radiation processes, and pointing towards the fact that these spectral-energy correlations are tightly reliant on the model-wise properties.

\end{abstract}

\keywords{Gamma-ray bursts (629); Astronomy data analysis (1858)}

\section{Introduction} \label{sec:intro}

Gamma-ray bursts (GRBs) are one of the most explosive events in the Universe. Two classes of GRBs have been identified in the CGRO/BATSE samples \citep{Kouveliotou1993} based on their duration\footnote{The time is taken to accumulate 90\% of the burst fluence starting at the 5\% fluence level.} $t_{90}$ with a separation line $t_{90}\approx 2$ s, long bursts (lGRBs) with a $t_{90}\gtrsim 2$ s and short bursts (sGRBs) with a $t_{90}\lesssim 2$ s. The two GRB populations invoke distinct physical progenitors: lGRBs formed by the massive-star collapse and sGRBs generated by the binary neutron star or neutron star-black hole merger. Different progenitors of the two GRB populations may lead to different observational properties (e.g., duration, total energy, spectral properties, and parameter correlations).

Correlation analysis plays an important role in the understanding of GRB physics as it provides a crucial clue to revealing their nature \citep[e.g.,][ and references therein]{Amati2002,Yonetoku2004,Liang2005,Amati2006,Dainotti2008,Yonetoku2010,Dainotti2008,Dainotti2010,Xu2012,Zhang2012,Heussaff2013,Dainotti2013,Dainotti2015,Wang2015NewAR,Dainotti2016,Dainotti2017,Dainotti2017a,Dainotti2018,Xu2021}. The two most widely discussed empirical correlations related to prompt emission mechanisms are the Amati \citep{Amati2002} and Yonetoku \citep{Yonetoku2004} relations, and both invoke the rest-frame peak energy $E_{\rm p,z}$=(1+z)$E_{\rm p}$ of the $\nu F_\nu$ prompt emission spectrum. The $E_{\rm p,z}$ strongly correlates with the GRB isotropic-bolometric-equivalent emission energy $E_{\gamma, \rm iso}$ at the cosmological rest frames, the so-called Amati relation ($E_{\rm p,z} \propto E^{0.5}_{\gamma, \rm iso}$), which was first discovered in \cite{Amati2002} using a very small burst sample (twelve GRBs with redshift estimates) detected by BeppoSAX, and subsequently confirmed by larger samples detected by other satellites (e.g., HETE II, Konus/WIND, INTEGRAL, {\it Swift}, and {\it Fermi}). These correlations were first established for lGRBs \citep{Amati2002}, and further research \citep[e.g,][]{Ghirlanda2009,Amati2010} has shown that sGRBs have comparable $E_{\rm p}-E_{\gamma, \rm iso}$ correlation features to long GRBs, but do not share the same $E_{\rm p}-E_{\gamma, \rm iso}$ correlation \citep[e.g.,][]{Zhang2012,Zhang2018}. This is because the spectra-energy properties between lGRBs and sGRBs are usually distinctly different \citep{Kouveliotou1993}, sGRBs are typically hard with a relatively high $E_{\rm p}$ \cite[e.g., ][]{Ghirlanda2004} while lGRBs are typically soft with a relatively low $E_{\rm p}$ \cite[e.g., ][]{Ghirlanda2009}. This may also be attributed to their short duration, as given the same $E_{\rm p,z}$, short bursts are typically less energetic \citep{Zhang2018book}.

The Amati correlation has been widely used as a powerful tool in understanding the nature and differences of GRBs for the following aspects: (i) studying the physics of jet structure and GRB unification scenarios \citep{Amati2006}; (ii) investigating the existence of different sub-classes of GRBs, such as distinguishing between the properties of long and short bursts \citep[e.g.,][]{Amati2002,Amati2006}, or diagnosing the classification properties of different pulses within a burst (Li et al. 2023, in prep.); (iii) as a cosmological tool applied to GRBs \citep[e.g.,][for a  review]{Wang2015NewAR}, for discussion on selection biases of the use of this and other prompt correlations see \cite{Dainotti2018}; (iv) it can be used to diagnose the radiation mechanism of GRBs, e.g., the synchrotron shock models \citep{Rees1994} or the photospheric emission models \citep{Rees2005,Peer2006a}, since these observational correlations can be reproduced by the photospheric emission models, however, they may be difficult to reproduce within the framework of the synchrotron shock models. Another related correlation, i.e., the correlation between the peak energy $E_{\rm p,z}$ and peak luminosity $L_{\rm p,iso}$, known as the Yonetoku relation, was discovered in \cite{Yonetoku2004}, and was used as the standard candle to estimate the redshift of the 689 GRBs with no known distances in the BATSE catalog. Unlike the Amati correlation, several previous studies \cite[e.g.,][]{Zhang2009b,Ghirlanda2009,Guiriec2013} revealed that short and long GRBs are no longer well distinguished in the $E_{\rm p,z}-L_{\gamma, \rm p, iso}$ plane, suggesting similar radiation processes.

Before {\it Fermi}, the observations covered a relatively narrower window into the energy (e.g., {\it Swift}-BAT; 15-350 keV, \citealt{Barthelmy2005}). Usually, determining $E_{\rm p}$ from the spectral analysis is a difficult task. For instance, the BAT is a narrow-band (15-150 keV) instrument, so constraining $E_{\rm p}$ straight from the Band-function spectral fit is frequently challenging due to the fact that in some cases $E_{\rm p}$ typically beyond the passband of the instrument. It is therefore likely that $E_{\rm p}$ cannot be measured accurately, at least for a significant fraction of the {\it Swift}-detected bursts.

Having both the Gamma-ray Burst Monitor (GBM; 8 KeV-40 MeV, \citealt{Meegan2009}), and the Large Area Telescope (LAT; 20 MeV-300 GeV, \citealt{Atwood2009}) onborad the {\it Fermi} Gamma-ray Space Telescope, it provides unprecedented spectral coverage of up to seven orders of magnitude in energy, and making it possible to fully assess all the current GRB spectral models \citep[e.g.,][]{Abdo2009a,Ryde2010,Axelsson2012,Guiriec2015a,Li2019a,Li2019c,Li2021b}. By May 2022, {\it Fermi} had completed 13 years of operation, in which at least 3000 GRBs had been observed, and contained at least 153 bursts with known redshifts, making it possible to study the GRB spectral-energy correlations with a large {\it Fermi} burst sample. There are several time-integrated and time-resolved spectral parameter catalogs of GRBs in the literature \citep[e.g.,][]{vonKienlin2014,Gruber2014,vonKienlin2020,NarayanaBhat2016,Yu2016} based on the {\it Fermi} observations, but the majority of them focus on the parameter properties (e.g., parameter distributions and parameter correlations) in the GRB observer frame. A few studies \citep[e.g.,][]{Poolakkil2021} based on the GBM data catalog have also presented their results in the GRB rest frame, however no relevant scientific research on the spectral parameters has yet been involved. Moreover, the frequentist method is the foundation for the majority of the earlier GBM catalogs \citep[e.g.,][]{vonKienlin2014,Gruber2014,vonKienlin2020,NarayanaBhat2016,Yu2016}. However, numerous recent studies \citep[e.g.,][]{Yu2019,Li2021b} have used a fully Bayesian approach. In light of this, it is important to perform independent analyses with a third party other than the official \emph{Fermi}/GBM team and to revisit these correlations based on the {\it Fermi} observations with a large burst sample. Specifically, the establishment of a spectral-energy correlation for a full sGRB sample is required. Moreover, previous studies \citep[e.g.,][]{Amati2002,Yonetoku2004,Ghirlanda2007,Amati2008,Nava2012,Qin2013,Minaev2020} have not given much attention to the properties of GRB pulses. In the framework of the standard fireball shock model, each pulse on the lightcurve relates to the emission formed by the collision of two fast and slow relativistic shells ejected from the central engine as a result of the shock-waves. Different pulses reflect different properties from the central engine and possibly also from the progenitor, such as some characteristic internal time and energy that is required to produce a pulse \citep{Li2021b}. Therefore, revisiting the GRB spectral-energy correlations based on their pulse properties is of great interest. In addition, it is also important to note that the majority of previous studies \citep[e.g.,][]{Nava2012,Qin2013} have directly used the spectral parameters of a single spectral model (e.g., Band) provided by the satellite online catalog (e.g., \emph{Fermi}/GBM). However, decades of observations have revealed that GRBs have diverse spectral properties that a single spectral model (e.g., Band-alone) cannot accurately characterize all the spectral shapes. For example, some GRB spectra can be well-fitted with a single non-thermal spectral component such as the Band-like component \citep[e.g., 080916C,][]{Abdo2009a}\footnote{Recent studies \citep{Guiriec2015a,Vereshchagin2022} have shown that in order to obtain an acceptable fit to the spectral data of GRB 080916C, a thermal component needs to be added during the initial prompt emission of the burst.}, but some other GRBs may require a dominant thermal component in order to obtain an acceptable fit \citep[e.g., 090902B,][]{Ryde2010}, and even some bursts exhibit a hybrid spectrum \citep[e.g., 110721A,][]{Iyyani2013}, i.e., composited with a non-thermal component and a thermal component within a single GRB simultaneously. Moreover, the $E_{\rm p}$ obtained from the non-thermal spectral fit is clearly less than that from the best fit in the hybrid spectrum if the spectral component is fully attributed to the non-thermal component \citep{Li2019c}. Furthermore, a recent study \citep{Li2022} suggests that inconsistent peaks (both $\alpha$ and $E_{\rm p}$) of the spectral parameter distribution have been found between Band-like bursts and CPL-like bursts. In particular, the derived spectral parameters deviated significantly in the ``Band (preferred)-to-CPL (misused)" case, but do not occur in the ``CPL (preferred)-to-Band (misused)" case\footnote{Here the ``Band (preferred)-to-CPL (misused)" case means that if a spectrum is statistically preferentially fitted to the Band, applying CPL to derive the spectral parameters, and vice versa the ``CPL (preferred)-to-Band (misused)" case.}.

Following these lines of argument, we will consider the following improvements in our investigations over earlier studies. 
(i) By using the broad spectral coverage of the {\it Fermi} data, the $E_{\rm p}$ of the prompt emission spectrum could accurately be measured for the majority of bursts.
(ii) all of the $E_{\rm p}$ in our tasks is obtained from the best-model fits by performing detailed spectral analysis and model comparisons between various GRB spectral models and their hybrid versions. 
(iii) Using a ``clean" sample of well-defined single-pulse GRBs and well-separated multi-pulse GRBs (Li et al. 2022, in prep.) to revisit the GRB spectral-energy correlations can in principle more directly reflect some internal properties of the central engine and predecessor stars. We will report our results in a series of papers using the {\it Fermi}-detected burst samples with distinct model and pulse properties. As the first paper in the series, we collect a complete GRB sample detected by {\it Fermi} with a measured redshift, and use the sample to revisit the GRB spectral-energy correlations, paying special attention to the Amati and Yonetoku relations. This effort is dedicated to achieving a ``clean" model-based GRB spectral-energy correlation analysis. 

The paper is organized as follows. The methodology are presented in Section 2. The results are summarized in Section 3. The discussions and conclusions are presented in Section 4 and Section 5, respectively. Throughout the paper, the standard $\Lambda$-CDM cosmology with the parameters $H_{0}= 67.4$ ${\rm km s^{-1}}$ ${\rm Mpc^{-1}}$, $\Omega_{M}=0.315$, and $\Omega_{\Lambda}=0.685$ are adopted \citep{PlanckCollaboration2018}.

\section{Methodology} \label{sec:data}

In order to perform the model-wise analysis of the spectral-energy correlations of GRBs proposed in this paper, a nearly complete GRB sample detected by {\it Fermi} with a measured redshift was created via a dedicated search from the NASA/HEASARC database\footnote{\url{https://heasarc.gsfc.nasa.gov/W3Browse/fermi/fermigbrst.html}} from July 2008 to May 2022, consisting of 153 such GRBs.  
Following the traditional GRB classification scheme according to their duration, 17 GRBs belong to short bursts ($t_{90}\lesssim 2$ s) whereas 136 GRBs belong to long bursts ($t_{90}\gtrsim 2$ s)\footnote{We notice that there are several bursts where we suspect that the duration reported by the GBM team may not be reliable. For instance, GRB 140506A880, GRB 191011192.}. Interestingly, with a time-dilation factor of 1/(1+z) corrected to the rest frame, the duration of six more GRBs (GRB 090423, GRB 110731A, GRB 130612A, GRB 140808A, GRB 141004A, and GRB 210610A) satisfy $t_{90}/(1+z)\lesssim2$ s. In order to obtain $E_{\rm p}$, $E_{\gamma, \rm iso}$, and $L_{\rm p, iso}$, our data procedure invokes the following steps, which we briefly introduce here.

\begin{enumerate}

\item Using {\tt 3ML} ({\tt the Multi-Mission Maximum Likelihood Framework}, see \citealt{Vianello2015}), and following the standard practices \citep{Li2019b,Li2019c,Yu2019,Burgess2019,Li2020,Li2021a,Li2021b} provided by the {\it Fermi} team, including the selection of detectors, sources, and background intervals, we performed a detailed spectral analysis for each individual burst in our initial sample. In order to ensure consistency of the results across various algorithms, we utilize both the maximum-likelihood estimation (MLE) and a fully Bayesian analysis plus Markov Chain Monte Carlo (MCMC) algorithms to explore their best parameter space.
	
\item For a given burst in our target sample, we first attempted using the GRB model, known as the Band function \citep{Band1993}, to fit its time-integrated spectral data. The photon number spectrum of Band is defined as
\begin{eqnarray}
N_{\rm Band}(E)=A \left\{ \begin{array}{ll}
(\frac{E}{E_{\rm piv}})^{\alpha} \rm exp (-\frac{{\it E}}{{\it E_{\rm 0}}}), & E < (\alpha-\beta)E_{\rm 0}  \\
\lbrack\frac{(\alpha-\beta)E_{\rm 0}}{E_{\rm piv}}\rbrack^{(\alpha-\beta)} \rm exp(\beta-\alpha)(\frac{{\it E}}{{\it E_{\rm piv}}})^{\beta}, & E\ge (\alpha-\beta)E_{0}\\
\end{array} \right.
\label{eq:Band} 
\end{eqnarray}
where $A$ is the normalization factor in units of ph cm$^{-2}$keV$^{-1}$s$^{-1}$, $E_{\rm piv}$ is the pivot energy always fixed at 100 keV, $E_{0}$ is the break energy correlated with the peak energy of $\nu F_{\nu}$ spectrum (assuming $\beta<-2$) by $E_{\rm p}=(2+\alpha)E_{\rm 0}$, $\alpha$ and $\beta$ are the low-energy and high-energy asymptotic power-law photon indices, respectively. There are two steps. (1) If all the model parameters from the Band fit are well-constrained, we then attempt to add a BB component to the Band. If an acceptable fit can still be obtained, we then obtain the $E_{\rm p}$ from the Band+BB fit. Otherwise, the Band-alone fit provides the $E_{\rm p}$. The BB emission can be modified by Planck spectrum, which is given by the photon flux
\begin{equation}
N_{\rm BB}(E,t)=A(t)\frac{E^{2}}{\rm exp\lbrack \frac{{\it E}}{{\it kT}(t)}\rbrack-1},
\end{equation}
where $A$ is the normalization, $T$ is the temperature, and $k$ is the Boltzmann constant. (2) Alternatively, if the model parameters from the Band fit are not well-constrained or Band-$\beta$ is poorly constrained (have fairly large values and large uncertainties), we then try the CPL model (a power law with an exponential tail) to refit the same spectral data and possibly obtain equally good fits for $\alpha$ and cut-off energy $E_{\rm c}$, and obtain the peak energy $E_{\rm p}$ of the $\nu F_\nu$-spectrum through\footnote{If the model parameters from the CPL model fit are not yet well-constrained, there may be two possibilities. (i) The lack of source photons in the analyzed bursts (e.g., $S<$ 20), so that the spectral fit can not be well-determined; or (ii) the source photons in the analyzed bursts are sufficient (e.g., $S>$ 20), but the model that best characterizes the spectral shape indeed is a simpler model than the CPL function (e.g., the simple power-law model).} $E_{\rm p}$ =(2+$\alpha$)$E_{\rm c}$. The CPL (COMP) function is given by 
\begin{equation}
N_{\rm CPL}(E) =A \left(\frac{E}{E_{\rm piv}}\right)^{\alpha}\rm exp(-\frac{\it E}{\it E_{c}}).
\label{CPL}
\end{equation}
Repeat Step (1), we may obtain $E_{\rm p}$ from the CPL+BB fit in some cases. As a result, $E_{\rm p}$ in our analysis can be obtained from two single spectral models (Band-alone and CPL-alone) and two hybrid spectral models (Band+BB and CPL+BB). To evaluate different spectral models and select the preferred one, we adopted both the Akaike Information Criterion (AIC; \citealt{Akaike1974}) and the Bayesian Information Criterion (BIC; \citealt{Schwarz1978}). This is due to the fact that the BIC is recommended for nested models (e.g., Band against Band+BB) while the AIC is favored for models that are not nested (such as Band versus CPL). The preferred model is the one that provides the lowest AIC and BIC scores. It should also be noted that in some cases, even if we obtain the lowest AIC and BIC scores, some of the model parameters found in the ``preferred" model cannot be constrained. Therefore, it is likely that we have found only the local minimum of the likelihood function rather than the global minimum. In this case, we need to reset the initial model parameters and repeat the fit until all model parameters are constrained and the minimum AIC and BIC scores are obtained.

\item Through Step 2, the $E_{\rm p}$ of the $\nu F_\nu$ prompt emission spectrum can be obtained from our refined spectral analysis, and its cosmological properties can be computed by 
\begin{equation}
E_{\rm p,z}=E_{\rm p}(1+z). 
\end{equation}
In addition, the energy flux $F_{\gamma}$ (erg cm$^{-2}$s$^{-1}$) can be also obtained from spectral parameters, with a $k$-correction ($k_c$, 1-10$^{4}$ keV) applied. In order to use the $E_{\rm p,z}$-$L_{\rm p}$ relation, a bolometric luminosity in a common cosmological rest-frame energy band (1-$10^{4}$ keV) is needed. It can be obtained by using the spectral parameters to conduct a $k$-correction extrapolating the observed energy band to 1-$10^{4}$ keV. For a given burst, the $k$-correction factor ($k_c$) can be derived using the following procedure. The observed flux $F^{\rm obs}$ (erg cm$^{-2}$ s$^{-1}$), in a fixed detector energy bandwidth [$e_1$, $e_2$] (for instance, for the {\it Fermi}-GBM observation, $e_1$=8 keV, $e_2$=40 MeV), can be written as: 
\begin{equation}
F^{obs}_{[e_1, e_2]} = \int_{e_1}^{e_2} EN(E)dE,
\label{eq:fluxIntegration}
\end{equation}
where $E$ is in units of keV, and $N(E)$ is a GRB photon number spectrum. The total luminosity emitted in the bandwidth [$e_1$,$e_2$], defined in the cosmological rest-frame, is given by:
\begin{equation}
    L_{[e_1(1+z), e_2(1+z)]} = 4 \pi D_L^2(z) F^{\rm obs}_{[e_1, e_2]}, 
\label{eq:CorrectedLuminosity}
\end{equation}
which $D_L(z)$ is the luminosity distance. To express the luminosity $L$ in the cosmological rest-frame energy band, [$E_1$=1 \rm{keV}, $E_2$ =10$^{4}$ \rm{keV}], common to all sources, the Eq.(\ref{eq:CorrectedLuminosity}) can be rewritten as:
\begin{align}
   L_{[E_1, E_2]}  &= 4 \pi D_L^2 F^{\rm obs}_
   {\left[\frac{E_1}{1+z},\frac{E_2}{1+z}\right]} =4 \pi D_L^2 k[e_1,e_2,E_1,E_2,z] F^{\rm obs}_{[e_1,e_2]}, 
\label{eq:???}
\end{align}
where the $k$-correction factor, $k_c$, is therefore defined as:
\begin{align}
k_c=k[e_1,e_2,E_1,E_2,z] =
& \frac{F^{\rm obs}_{\left[\frac{E_1}{1+z};\frac{E_2}{1+z}\right]}}{ F^{\rm obs}_{[e_1,e_2]}} =\frac{\int_{E_1/(1+z)}^{E_2/(1+z)} EN(E)dE}{\int_{e_1}^{e_2} EN(E)dE}, 
\end{align}
with $k_{\rm c}$, $F_{\gamma}$, and redshift measurement, one can estimate the peak isotropic-equivalent luminosity as
\begin{equation}
L_{\rm p, iso}=4\pi D^{2}_{L} F_{\rm p,\gamma} k_c, 
\end{equation}
where $D_{L}$ is the luminosity distance.
Therefore, the fluence $S_{\gamma}$ (erg cm$^{-2}$) during the source interval ($\Delta T_{\rm src}$) can be yielded by $S_{\gamma}$=$F_{\gamma}$$\Delta T_{\rm src}$.
With $S_{\gamma}$, and redshift measurement, one can estimate the isotropic-equivalent energy releases in $\gamma$-ray band
\begin{equation}
E_{\gamma,\rm iso}=\frac{4\pi d^{2}_{L} S_{\gamma} k_c}{(1+z)}.
\end{equation}
\end{enumerate}

\section{Results} \label{sec:result}

\subsection{Distributions of $k_{\rm c}$, $T_{90}$ ($T_{90,z}$), $S_{\gamma}$, $E_{\rm p,z}$, $E_{\gamma,\rm iso}$, and $L_{\rm p, iso}$} 

\begin{figure*}[ht!]
\includegraphics[width=1.\hsize,clip]{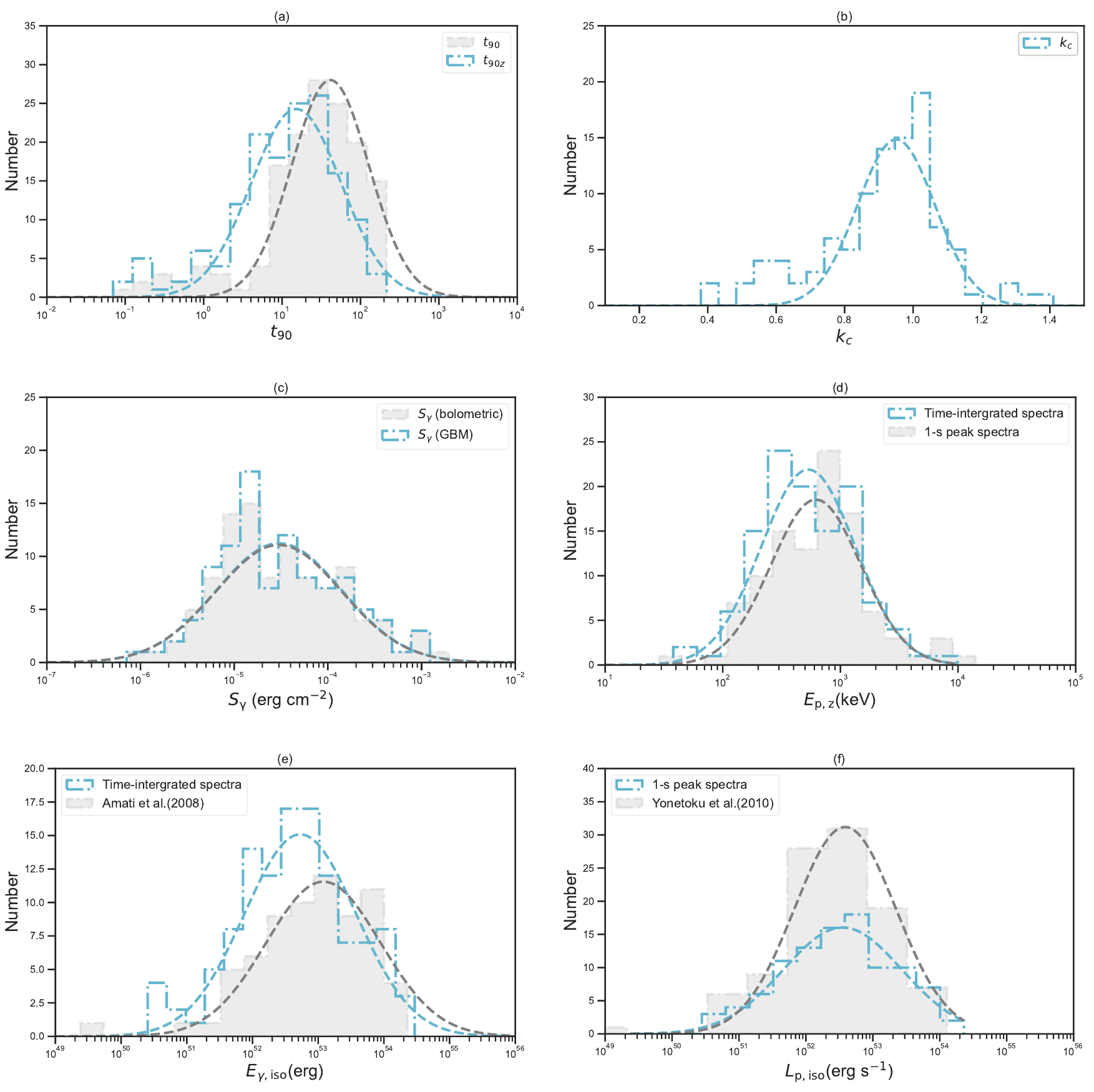}
\caption{Global parameter distributions of (a) $T_{90}$ (cyan dotted dashed line) and $T_{90,z}$ (grey shaded area), (b) $k_{\rm c}$, (c) $S^{\rm bolometric}_{\gamma}$ (cyan dotted dashed line) and $S^{\rm GBM}_{\gamma}$ (grey shaded area), (d) $E_{\rm p,z}$ between time-intergrated spectra (cyan dotted dashed line) and 1-s peak spectra (grey shaded area), (e) $E_{\gamma,\rm iso}$ between time-intergrated spectra (cyan dotted dashed line) and the same defined in \cite{Amati2008} (grey shaded area), and (f) $L_{\gamma,\rm iso}$ between 1-s peak spectra (dotted dashed line) and \cite{Yonetoku2010} (grey shaded area). The (cyan and grey) dashed lines represent their best Gaussian fits.}
\label{fig:dis}
\end{figure*}

We show the distributions of $k_{\rm c}$, $T_{90}$ ($T_{90,z}$), $S_{\gamma}$, $E_{\rm p,z}$, $E_{\gamma,\rm iso}$, and $L_{\gamma,\rm iso}$ in Figure \ref{fig:dis} for our complete sample described in Section \ref{sec:data}. Figure \ref{fig:dis}a depicts the distributions of $t_{90}$ for the cases both in the observed frame (cyan line) and rest-frame (shadow grey region), and one can see that they do not share the same distribution. With a time-dilation factor 1/(1+z) corrected, the peak of the distribution of $t_{90,z}$ at the cosmological rest-frame is smaller than that of the observed frame (see Table \ref{tab:Gaussian}). The $k$-correction factor, $k_{\rm c}$, is calculated as the flux ratio between the 1-10$^{4}$ keV and GBM energy bands (8 keV-40 MeV) based on the time-integrated spectra of each burst in the sample. The distribution can be fitted with Gaussian functions ($\mathcal{N}=\mu \pm \sigma$), where $\mu$ is the average value and $\sigma$ is the corresponding standard deviation. The best Gaussian fit for the distribution of $k_{\rm c}$ gives $k_{\rm c}=0.95\pm0.11$ (Figure \ref{fig:dis}b). This result ($k_{\rm c}\approx$1) is due to the GBM energy band being comparable to the $k$-correction energy band that we used. Similar results can also be found in Figure \ref{fig:dis}c, where we present the distributions of $S_{\gamma}$. The $S_{\gamma}$ obtained in the GBM band is shown by the cyan line while that of in the bolometric (1-10$^{4}$ keV) energy band is overlaid in gray. One can see that both share a similar distribution. The spectral parameters ($E_{\rm p,z}$, $E_{\gamma,\rm iso}$, and $L_{\gamma,\rm iso}$) are derived from the model-wise spectral analysis (see Section \ref{sec:data}). As presented in Section \ref{sec:data}, $E_{\rm p,z}$  can be obtained from both the time-integrated (see dashed line in Figure \ref{fig:dis}d) and (1-s) peak (see shadow grey region in Figure \ref{fig:dis}d) spectral analyses. However, $E_{\gamma,\rm iso}$ is only obtained from the time-integrated spectral analysis (see dashed line in Figure \ref{fig:dis}e) whereas $L_{\gamma,\rm iso}$ is only obtained from the (1-s) peak spectral analysis (see dashed line in Figure \ref{fig:dis}f). Detailed information about the best Gaussian fit of each distribution is presented in Table \ref{tab:Gaussian}, along with the corresponding average values and standard deviations.

\subsection{The model-wise $E_{\rm p,z}-E_{\gamma, \rm iso}$ (Amati) Correlation} \label{sec:Amati}

With the refined spectral analysis described in Section \ref{sec:data}, one hundred and nine GRBs have well-measured their time-intergreated $E_{\rm p}$, whereas thirty-eight GRBs have a low statistical significance\footnote{Note that there is a peculiar event (GRB 150727A). The event has a high statistical significance and a well-measured $E_{\rm p}$. However, background photons cannot be properly subtracted, resulting in somewhat anomalous results. The spectral parameters ($\alpha$=-1.12$\pm$0.13, $E_{\rm c}$=210$^{+75}_{-55}$ keV) obtained from the CPL model (better than other models) fit, with reshift at z=0.313, we obtain $E_{\rm p,z}$=242$^{+93}_{-72}$ keV and $E_{\gamma,\rm iso}$=(0.5$^{+0.2}_{-0.1}$) $\times$ 10$^{52}$ erg for this event.}, resulting in poorly determined spectral fits and hence unmeasured $E_{\rm p}$ (see Column 8 in Table \ref{tab:global}). The 109 GRBs with well-measured $E_{\rm p}$ provide a well-defined sample for studying the GRB time-integrated $E_{\rm p,z}-E_{\gamma, \rm iso}$ (Amati) correlation. Based on the time-integrated spectral analysis, we independently present all the fitted parameters (unless the model parameters cannot be well-constrained for some bursts) using two individual (standard) models (CPL and Band) in Table \ref{tab:Integrated}. The information given in Table \ref{tab:Integrated} include GRB name (Column 1); The selected source interval $T_{\rm start} \sim T_{\rm stop}$ (Column 2); The corresponding significance $S$; The best-fit parameters for the CPL model (normalization $K$, low-energy power-law index $\alpha$, and cutoff energy $E_{\rm c}$ of the $\nu F_{\nu}$ spectrum) are listed in Columns (4-6); as well as the corresponding likelihood, AIC, and BIC; The derived rest-frame peak energy $E_{\rm p,z}$ and the isotropic-bolometric-equivalent $\gamma$-ray emission energy $E_{\rm \gamma,iso}$ are listed in Columns 8-9. The best-fit parameters for the Band model (normalization $K$, low-energy power-law index $\alpha$, peak energy $E_{\rm p}$ of the $\nu F_{\nu}$ spectrum, and high-energy power-law index $\beta$) are listed in Columns (10-13), as well as the corresponding likelihood, AIC, and BIC; The derived $E_{\rm p,z}$ and $E_{\rm \gamma,iso}$ are listed in Columns (15-16); The difference of AIC between Band and CPL models, defined as $\Delta$AIC=AIC$_{\rm Band}$-AIC$_{\rm CPL}$, is listed in Column 17. By performing model-wise analysis following the Steps described in Section \ref{sec:data}, we classify these bursts into four groups. To summarize, our sample is composed as follows.
\begin{itemize}
\item Band-like bursts. This group includes 64 GRBs (3 sGRBs and 61 lGRBs) in which the spectral data can be well fitted by the Band-alone model (see Column (8) of Table \ref{tab:global} and  Column (17) of Table \ref{tab:Integrated}), and $E_{\rm p}$ can be directly obtained from the fits, being consistent with the previous studies \cite[e.g.,][]{Amati2002}. 
\item CPL-like bursts. This group consist of 45 GRBs (6 sGRBs and 39 lGRBs) that CPL-alone model can be well fitted the spectral data (see Column (8) of Table \ref{tab:global} and  Column (17) of Table \ref{tab:Integrated}), and $E_{\rm p}$ is computed through $E_{\rm p}$ =(2+$\alpha$)$E_{\rm c}$.
\item Band+BB-like bursts. This group includes 6 lGRBs (0 sGRBs and 6 lGRBs) in which the spectral data require an additional thermal component based on the Band component, namely, the Band plus a BB model (see the upper panel of Table \ref{tab:hybrid}), and $E_{\rm p}$ can be directly obtained from the Band component.
\item CPL+BB-like bursts. This group consist of 5 lGRBs (0 sGRBs and 5 lGRBs) that the spectral data still require an additional thermal component based on the CPL model, namely, CPL plus a BB component (see the upper panel of Table \ref{tab:hybrid}), and $E_{\rm p}$ is computed through $E_{\rm p}$ =(2+$\alpha$)$E_{\rm c}$ from the CPL component.
\end{itemize}

To measure the $E_{\rm p}$ of the prompt emission spectra, the Band model \citep{Band1993} has been used extensively in previous studies. Several recent statistical results \citep[e.g.,][]{Yu2019,Li2021b} based on {\it Fermi} observations show that the Band-like spectra dominate the time-integrated spectral properties, whereas the CPL-like spectra dominate the time-resolved spectral properties, regardless of whether it is a single-pulse \citep{Yu2019} or multi-pulse \citep{Li2021b} burst. These two spectral models are widely used in practically all GRB literature and are the canonical models of GRBs. Furthermore, the deviations in the derived spectral parameters as a result of their misuse have also been thoroughly investigated in a recent study \citep{Li2022}. Moreover, hybrid spectra \citep[e.g., 110721A,][]{Axelsson2012} can be observed in some other GRBs, and the $E_{\rm p}$ determined from the spectral fit deviates greatly from the intrinsic spectral shape if the spectral component is totally attributed to the non-thermal component \citep{Li2019c}.

In light of the above arguments, it will be fascinating to see whether GRB spectrum-energy correlations are affected by the spectral model chosen. We first investigate the model-based properties of the $E_{\rm p,z}-E_{\gamma, \rm iso}$ (Amati) correlation. Similar to the method described in \cite{Nava2012}, we modeled the distribution of data points for our lGRB sample by using a linear function in the $E_{\rm p,z}-E_{\gamma, \rm iso}$ logarithmic plane, and fitted the data using a nonlinear least-squares method using the Levenberg-Marquardt minimization algorithm \citep{Newville2016}. This option is motivated by the fact that there is no reason to assume either $E_{\rm p}$ or $E_{\gamma, \rm iso}$ (or $L_{\gamma, \rm iso}$) as an independent variable a priori, as well as by the high degree of dispersion in the data points. The slope and normalization errors are computed by fitting data points to their barycenters, which are uncorrelated. We estimate the Spearman's rank correlation coefficient and the associated chance probability for the samples, and provide these values in Table \ref{tab:Spearman}. To double-check our results, we also performed a correlation coefficient analysis using the Markov Chain Monte Carlo algorithm to evaluate the correlation between these parameters (see Appendix and Table \ref{tab:MCMC}).

In Figure \ref{fig:Amati}, we present the (preferred) spectral-model-based $E_{\rm p,z}-E_{\gamma, \rm iso}$ (Amati) correlation analysis for our sample. The data points with magenta, cyan, orange, and grey colors indicate the Band-like, CPL-like, Band+BB-like, and CPL+BB-like bursts, receptively. We employ the power-law model $E_{\rm p,z}=a\left(\frac{E_{\gamma, \rm iso}}{E^{\rm piv}_{\gamma,\rm iso}}\right)^{b}$ to fit the data, where $a$ is the is the normalization, $b$ is the power-law index, and $E^{\rm piv}_{\gamma,\rm iso}$ is the pivot energy fixed at 2$\times$ 10$^{52}$ erg. The fits are performed by using the python package \emph{lmfit} \citep{Newville2016} by applying a nonlinear least-squares method using the Levenberg-Marquardt minimization algorithm. The best power-law fit is shown by the grey line, while the shadow area represents the 2$\sigma$ error zone. The power-law model fitted to our Band-like lGRBs using the Spearman correlation analysis gives
\begin{eqnarray}\label{PLfitRest}
{\it E_{\rm p,z} {\rm /(keV)}}=(229\pm29)\left[\frac{E_{\gamma,\rm iso}{\rm /(erg)}}{E^{\rm piv}_{\gamma,\rm iso}}\right]^{(0.42\pm0.05)},
\end{eqnarray} 
at 2$\sigma$ confidence level, with the number of data points $N$=61, the Spearman's rank correlation coefficient of $R$=0.71, and a chance probability $p< 10^{-4}$. The power-law model fitted to our CPL-like lGRBs bursts gives
\begin{eqnarray}\label{PLfitRest}
{\it E_{\rm p,z} {\rm /(keV)}}=(341\pm65)\left[\frac{E_{\gamma,\rm iso}{\rm /(erg)}}{E^{\rm piv}_{\gamma,\rm iso}}\right]^{(0.25\pm0.12)},
\end{eqnarray} 
at 2$\sigma$ confidence level, with the number of data points $N$=39, the Spearman's rank correlation coefficient of $R$=0.11, and a chance probability 0.67. Moreover, we found three Band-like short bursts and six CPL-like short bursts (see Column (8) in Table \ref{tab:global} and Column (17) of Table \ref{tab:Integrated}). Using mixed samples (sGRBs and lGRBs), we also attempted to do a similar model-wise analysis and reported our results in Table \ref{tab:Spearman}. For the Band-like cases, due to the small sample size (3 events), adding the sGRBs sample does not significantly affect the results. However, for the CPL-like case, the correlation changes significantly when the sGRB sample is included (see Table \ref{tab:Spearman}). Also, we note that due to the fact that the sample size for the hybrid spectral events is small, we are not able do a relevant statistical analysis for these groups. A more detailed statistical analysis of hybrid spectra, will be possible in the future when \emph{Fermi} observations accumulate more such cases.

\begin{figure*}[ht!]
\includegraphics[width=1.\hsize,clip]{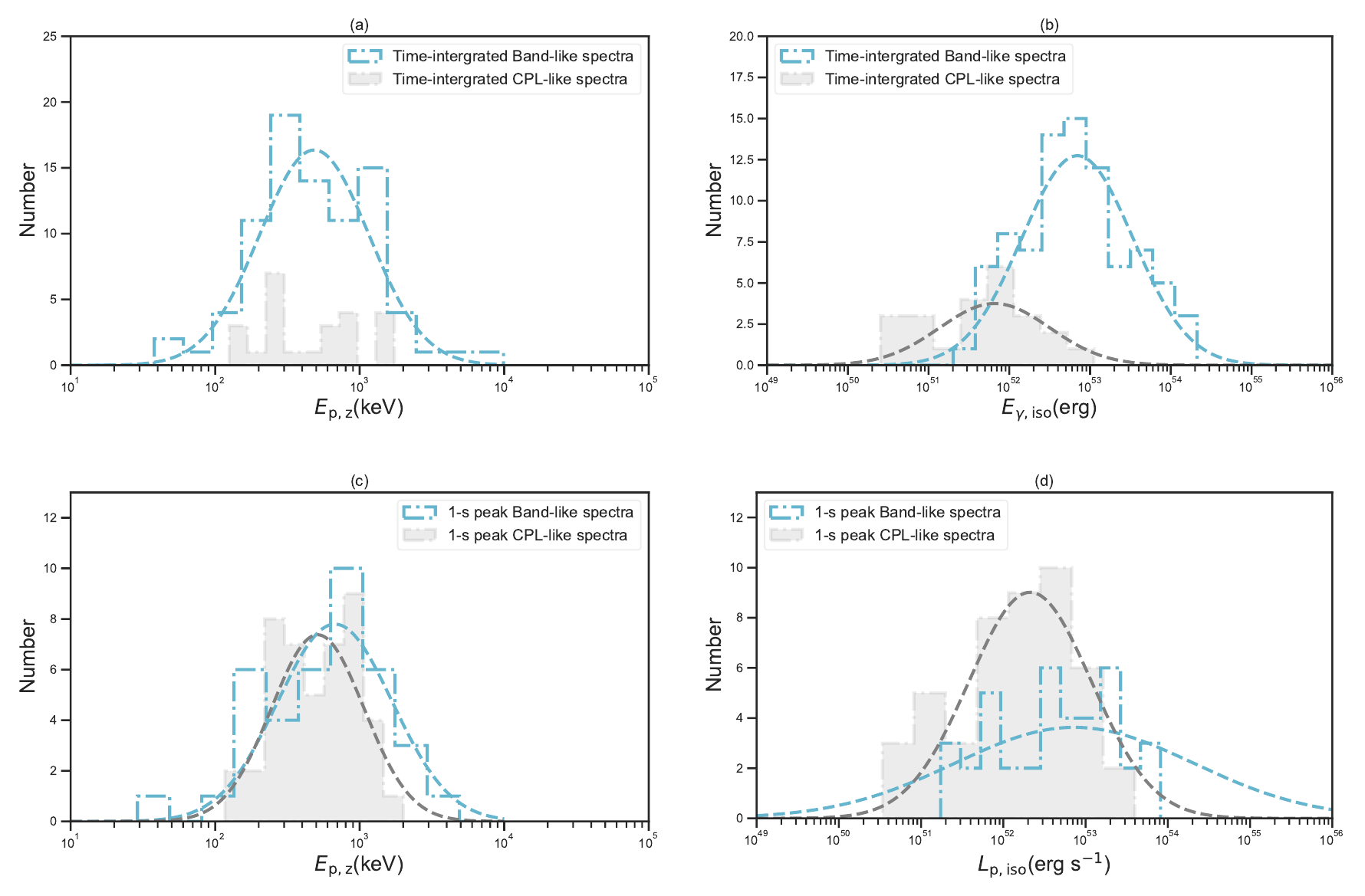}
\caption{Same as Figure \ref{fig:dis} but for the model-wise parameter distributions (Band versus CPL).}
\label{fig:dis_Band_CPL}
\end{figure*}

Our analysis shows that Band-like bursts and CPL-like busts may not have the same Amati correlation as shown in the $E_{\rm p}-E_{\gamma, \rm iso}$ plane (Figure \ref{fig:Amati}), despite the fact that CPL-like events have a more significant dispersion and that data points have a larger error bar, and therefore, the results may not be statistically significant. More interestingly, the Band-like spectral-based Amati correlation remains consistent with previous studies without distinguishing between the spectral models. The CPL-like spectral-based Amati correlation, on the other hand, is inconsistent with the findings in previous samples and with the Band-like bursts, and a shallower power-law index is found. Individual parameter distributions ($E_{\rm p}$ and $E_{\gamma, \rm iso}$) are shown in the upper panels of Figure \ref{fig:dis_Band_CPL}. Compared to the Band-like bursts, one can see that CPL-like bursts show a smaller $E_{\gamma, \rm iso}$ peak. However, CPL-like bursts and Band-like bursts appear to have similar peaks in the $E_{\rm p}$ distributions (Figure \ref{fig:dis_Band_CPL}a). As a result, Band-like bursts and CPL-like busts may not have the same Amati correlation. In addition, if the CPL-like spectra events can explain the outliers observed in the $E_{\rm p,z}$-$E_{\gamma,\rm iso}$ plane, they should dominate the low-$E_{\gamma,\rm iso}$ and high-$E_{\rm p,z}$ regions, as shown in Figure \ref{fig:Amati}.

\subsection{The model-wise $E_{\rm p,z}-L_{\rm p, iso}$ (Yonetoku) Correlation}

Investigating the spectral-model-based properties of the Yonetoku correlation is equally intriguing. In order to investigate the $E_{\rm p,z}-L_{\rm p, iso}$ (Yonetoku) correlation, we need to obtain both the time-averaged $E_{\rm p,z}$ and the peak luminosity $L_{\rm p, iso}$ according to the definition of the previous studies \citep[e.g.,][]{Yonetoku2010}. In order to obtain the peak luminosity $L_{\rm p, iso}$, we select the (1-s) peak spectrum of a given burst in our sample and repeat Steps 1-3 described in Section \ref{sec:data}. In order to precisely identify the 1-s peak energy spectrum, we apply the ``constant'' binning method with a time slice $\Delta t$=1 to the TTE (Time-Tagged Events) light curve of the brightest NaI detector. We then calculate the counts for each time bin based on the ``constant'' time binning method and pick up the one with the maximum count value. Fifty-one GRBs were too faint to determine the peak luminosity based on their spectral parameters. The remaining 102 GRBs that can determine their spectral parameters could be further used to study the $E_{\rm p,z}-L_{\rm p, iso}$ (Yonetoku) correlation. Combined with the time-integrated spectral analysis studied in Section \ref{sec:Amati}, a total of 92 GRBs were able to obtain both well-measured $E_{\rm p,z}$ from the time-integrated spectrum and peak luminosity $L_{\rm p, iso}$ from the 1-s peak spectrum, and can be divided into four groups based on their best-fit spectral models. Similarly, based on the 1-s peak spectral analysis, we also independently present all the fitted parameters from both CPL and Band models (Table \ref{tab:1speak}). The derived peak luminosity $L_{\rm p, iso}$ are listed in Column (8) and Column (14) for the CPL and Band models, respectively. The information remaining is the same as in Table \ref{tab:Integrated}.

\begin{itemize}
\item Band-like bursts. This group includes 55 GRBs (3 sGRBs and 52 lGRBs) in which the spectral data can be well fitted by the Band-alone model (see Column (8) of Table \ref{tab:global} and  Column (17) of Table \ref{tab:1speak}), and $E_{\rm p}$ can be directly obtained from the fit, being consistent with the previous studies \citep[e.g.,][]{Yonetoku2010}. 
\item CPL-like bursts. This group consist of 37 GRBs (2 sGRBs and 35 lGRBs) that CPL-alone model can be well fitted the spectral data (see Column (8) of Table \ref{tab:global} and  Column (17) of Table \ref{tab:1speak}), and $E_{\rm p}$ is computed through $E_{\rm p}$ =(2+$\alpha$)$E_{\rm c}$.
\item Band+BB-like bursts. This group includes 5 GRBs (0 sGRBs and 5 lGRBs) in which the spectral data require an additional thermal component based on the Band component, namely, the Band plus a BB model (see the lower panel of Table \ref{tab:hybrid}), and $E_{\rm p}$ can be directly obtained from the Band component.
\item CPL+BB-like bursts. This group consist of 8 lGRBs (0 sGRBs and 8 lGRBs) that the spectral data still require an additional thermal component based on the CPL model, namely, CPL plus a BB component (see the lower panel of Table \ref{tab:hybrid}), and $E_{\rm p}$ is computed through $E_{\rm p}$ =(2+$\alpha$)$E_{\rm c}$ from the CPL component.
\end{itemize}

Figure \ref{fig:Yonetoku} shows the spectral model-dependent $E_{\rm p}-L_{\rm p, iso}$ (Yonetoku) correlation for 102 GRBs with a well-measured $E_{\rm p}$ using 1-s peak spectrum. The symbols and colors are the same as in Figure \ref{fig:Amati}. Similarly, we employ the power-law model $E_{\rm p,z}=a\left(\frac{L_{\rm p,iso}}{L^{\rm piv}_{\rm p, iso}}\right)^{b}$ to fit the data, where $a$ is the is the normalization, $b$ is the power-law index, and $L^{\rm piv}_{\rm p, iso}$ is the pivot luminosity fixed at a typical value (see Table \ref{tab:Spearman}). We fit the Band-like 1-s peak spectra for our lGRB sample with the power-law model using the Spearman correlation analysis, and yield the following results
\begin{eqnarray}\label{PLfitRest}
{\it E_{\rm p,z} {\rm /(keV)}}=(307\pm34)\left[\frac{L_{\rm p, iso}{\rm /(erg.s^{-1})}}{L^{\rm piv}_{\rm p, iso}}\right]^{(0.34\pm0.04)},
\end{eqnarray} 
at 2$\sigma$ confidence level, with the number of data points $N$=53, the Spearman's rank correlation coefficient of $R$=0.83, and a chance probability $p< 10^{-4}$. Similar analysis is performed on the CPL-like lGRBs and come up with the following results
\begin{eqnarray}\label{PLfitRest}
{\it E_{\rm p,z} {\rm /(keV)}}=(454\pm55)\left[\frac{L_{\rm p, iso}{\rm /(erg.s^{-1})}}{L^{\rm piv}_{\rm p, iso}}\right]^{(0.40\pm0.08)},
\end{eqnarray} 
at 2$\sigma$ confidence level, with the number of data points $N$=35, the Spearman's rank correlation coefficient of $R$=0.69, and a chance probability $p< 10^{-4}$. 

Based on the time-integrated spectral analysis and the 1-s peak spectral analysis, we found that the CPL-like lGRBs do not fall on the Band-like lGRB Yonetoku correlation, having a shallower slope, as shown in the $E_{\rm p,z}-L_{\rm p, iso}$ plane (Figure \ref{fig:Yonetoku}). This result is inconsistent with the findings \cite[e.g.,][]{Amati2006,Zhang2009b,Ghirlanda2009,Guiriec2013} that short and long GRBs in the $E_{\rm p,z}-L_{\rm p, iso}$ plane are no longer well-separated. However, it is similar to the findings found in the $E_{\rm p,z}-E_{\gamma,\rm iso}$ plane in several previous studies that the short GRBs do not fall on the long GRB Amati relation \cite[e.g.,][]{Zhang2009b,Ghirlanda2009,Guiriec2013}. Moreover, we also performed a similar model-wise analysis by adding the sGRBs to the lGRB sample (see Table \ref{tab:Spearman}). Due to the small sample size of the sGRBs for both the Band-like and CPL-like cases, we do not find a significant difference between the lGRB sample and entire (mixed) sample (sGRBs+lGRBs). We discover that the Band-like events have a  larger dispersion than the CPL-like events, and that the power-law index is inconsistent with the CPL-like bursts with respect to their uncertainties. This may be due to the fact that the sample size of the Band-like event is not large enough to lead to missed low-luminosity events.

\subsection{Outliers in the $E_{\rm p,z}$-$E_{\gamma,\rm iso}$ Plane} 

Interestingly, we discovered three notable outliers (GRB 110721A, GRB 130702A, and GRB 140606B) that are located outside the 3$\sigma$ region of the lGRB Amati correlation, as shown in the $E_{\rm p,z}-E_{\gamma, \rm iso}$ plane, where all these events belong to the traditional classification of long-duration bursts (Figure \ref{fig:outlier}). More interestingly, the CPL model presents the best fittings to GRB 130702A and GRB 140606B while the remaining one (GRB 110721A) is a Band+BB-like burst. These results motivated us to investigate whether spectral modeling could affect the results of the Amati correlation and whether the selection of the spectral model could have contributed to the outliers in the $E_{\rm p,z}$-$E_{\gamma,\rm iso}$ plane. Apart from these outliers, several more events that either involve Band+BB-like spectra or CPL+BB-like spectra are also interesting to include. Together, these events are particularly useful for testing whether the outliers observed in the $E_{\rm p,z}-E_{\gamma, \rm iso}$ plane arise from the applied spectral model selection. We next compare the $E_{\rm p}$, $E_{\gamma, \rm iso}$, and $E_{\rm p}-E_{\gamma, \rm iso}$ correlations between the preferred model and the Band model by selecting several outliers from CPL-like, Band+BB-like, and CPL+BB-like bursts, as shown in Figure \ref{fig:outlier}. The CPL model is statistically preferred for GRB 101213A, and GRB 140606B with respect to values of $\Delta$AIC=2.0 and 1.6 (see Column 17 of Table \ref{tab:Integrated}), respectively. Moreover, we discovered that $E_{\gamma,\rm iso}$ did not change when alternative spectral models were used, since energy fluence (erg cm $^{-2}$) integrated from energy flux (erg cm$^{-2}$s$^{-1}$) between the same time period and energy range (e.g., 1-10$^{4}$ keV) based on various spectral models varied very little \citep{Li2019c,Li2022}. For the other two cases (GRB 101213A and GRB 140606B), neither $E_{\rm p,z}$ nor $E_{\gamma,\rm iso}$ altered after Band model application, which corresponds to a recent study \citep{Li2022}, where we found that the derived spectral parameters deviated significantly in the ``Band (preferred)-to-CPL (misused)" case, but did not occur in the ``CPL (preferred)-to-Band (misused)" case \citep{Li2022}. Both the Band+BB-like and CPL+BB-like bursts invoke a hybrid spectrum, with a subdominant thermal component occupying the left shoulder (blow $E_{\rm p}$) of the Band or CPL component. For the Band+BB-like bursts, the Band+BB model is statistically preferred for GRB 110721A, GRB 150314A, and GRB 150403A with respect to values of $\Delta$BIC=-7.9, -14.5, and -115.6 (see Column 14 of Table \ref{tab:BB}), respectively. One notable outlier in the $E_{\rm p}-E_{\gamma, \rm iso}$ plane is GRB 110721A. While the derived $E_{\rm p}$ from the Band greatly differs from that of the Band+BB, it still does not fall back to the region dominated by the Amati relation. The $E_{\rm p}$ of GRB 150403A changed moderately, but neither Band+BB nor Band did not exhibit a significant outlier in the $E_{\rm p,z}-E_{\gamma, \rm iso}$ plane. The last case is GRB 150314A, which has a negligible alter in $E_{\rm p}$ and falls well in the $3\sigma$ region of the lGRB $E_{\rm p}-E_{\gamma, \rm iso}$ correlation. For the CPL+BB-like bursts, the CPL+BB model is statistically preferred for GRB 131231A, and GRB 181020A with respect to values of $\Delta$BIC=-331.2, and -3.9, respectively. Both cases show moderate alterations of $E_{\rm p}$ when the Band model is reused, and neither case has a substantial outlier in the $E_{\rm p,z}-E_{\gamma, \rm iso}$ plane.

Several possibilities have been proposed by some authors to explain the outliers of the Amati correlation holding for long cosmological GRBs \citep{Amati2002}. For instance, (i) due to viewing angle effects with off-axis scenarios; (ii) the existence of a class of nearby and intrinsically faint GRBs with different properties with respect to ``standard" GRBs; (iii) several earlier studies \citep[e.g.,][]{Amati2006} also point out that the Amati correlation can be used as a pseudo-redshift estimator for testing possible selection effects of GRBs with unknown redshifts. We, therefore, briefly discuss here the possible reasons for the three outliers (GRB 110721A, GRB 130702A, and GRB 140606B) observed in the $E_{\rm p,z}$-$E_{\gamma,\rm iso}$ plane of our lGRB sample. Interestingly, \cite{Berger2011} suggested two possible redshifts ($z$ = 0.382 or $z$ = 3.512) for GRB 110721A based on a candidate optical counterpart reported in \cite{Greiner2011}, with the former being preferred. To investigate whether the outlier (GRB 110721A) is caused by a pseudo-redshift, we conducted the following tests. By performing the model comparisons, we found that the Band+BB model superiors other single (e.g., Band) and hybrid (e.g., CPL+BB) models. The rest-frame peak energy $E_{\rm p,z}$=(1+z)$E_{\rm p}$, as well as isotropic total energy $E_{\gamma,\rm iso}=4\pi d^{2}_{L} S_{\gamma} k_c/(1+z)$, can be calculated using the spectral parameters obtained from the Band+BB model fit (Table \ref{tab:BB}). We then plot the evolution of the redshift from 0.01 to 10 in the $E_{\rm p,z}$-$E_{\gamma,\rm iso}$ plane for these outliers (see dashed lines in Figure \ref{fig:outlier}). To directly compare the possible values of redshift for the burst falling in the $E_{\rm p,z}$-$E_{\gamma,\rm iso}$ region with the possible values of redshift in the non-$E_{\rm p,z}$-$E_{\gamma,\rm iso}$ region, we use gray, black, and red to denote the low (from 0.01 to 1), intermediate (from 1 to 3), and high redshift (from 3 to 10) regions, respectively. Conclusively, our analysis indicates that $z$ = 3.512 could be a preferred redshift candidate for maintaining the validity of the $E_{\rm p,z}$-$E_{\gamma,\rm iso}$ correlation. However, the redshifts of both GRB 130702A \citep{Singer2013} and GRB 140606B \citep{Perley2014GCN} have been firmly measured. As a result, the method applied to GRB 110721A may not be applicable to GRB 130702A and GRB 140606B. For GRB 130702A, our refined spectral analysis suggests that the CPL model is a superior model compared to other models. With a redshift at $z$=0.145 (it is at a fairly close distance) and the spectral parameters obtained from the best fitting, where $E_{\rm p}$ =(2+$\alpha$)$E_{\rm c}$, and one can calculate its isotropic total energy as $E_{\gamma,\rm iso}$=$2.9^{+3.1}_{-1.4}\times 10^{50}$ erg. The unique feature of the event is its extremely low isotropic bolometric emission energy. Specifically, the presence of a bright supernova associated with GRB 130702A \citep{DElia2015} implies that it is a nearby event. GRB 140606B is another event similar to GRB 130702A. Its spectroscopically associated type Ic-BL SN was reported in \cite{Cano2015}. The time-integrated spectrum can be fitted by the CPL model with $\alpha$=-1.19$\pm$0.05 and $E_{\rm c}$=532$^{+141}_{-111}$ keV. With a redshift measured (z=0.384), the rest-frame peak energy ($E_{\rm p,z}=597^{+162}_{-130}$ keV) and isotropic bolometric emission energy ($E_{\gamma,\rm iso}=2.9^{+0.7}_{-0.5} \times 10^{51}$ erg) can be calculated. This event is thus located outside the 3$\sigma$ region of the lGRB Amati correlation (see Figure \ref{fig:outlier}). 

\section{Discussions} \label{sec:dis}

\subsection{Comparison with Previous Samples} 

Our sample selection criteria differ from previous sample studies \citep[e.g.,][]{Amati2008,Yonetoku2010} in several major ways. (i) In order to fully evaluate the  various current GRB spectral models, we focus on the bursts observed by {\it Fermi}-GBM. (ii) In order to allow a ``clean" study of the pulse properties of the energy-spectral correlations, we focus on a sample of well-defined single-pulse GRBs and well-separated multi-pulse GRBs. (iii) In order to investigate the spectral model-dependent properties of the Amati and Yonetoku relations, the $E_{\rm p}$ for each burst in our sample was obtained using a preferred spectral model by examining several frequently used spectral models. Based on these criteria, our sample selection may have a bias. Thus, an interesting question is whether the spectral-energy parameters we obtained differ from the previous samples in terms of statistical distributions. Here, we address the question of whether or not the spectral-energy properties of our sample are similar to those of the previous samples.

We compare the distributions of $E_{\rm p,z}$ (Fig.\ref{fig:dis}d), $E_{\gamma,\rm iso}$ (Fig.\ref{fig:dis}e), and $L_{\rm p, iso}$ (Fig.\ref{fig:dis}f), between our sample and the samples in \cite{Amati2008} and \cite{Yonetoku2010}. Our sample is shown by the cyan dashed line while the sample defined in \cite{Amati2008} or \cite{Yonetoku2010} is displayed by the grey shaded area. The best Gaussian fit for each distribution, including the corresponding average values and standard deviations are summarized in Table \ref{tab:Gaussian}. We find that both $E_{\rm p,z}$ and $E_{\gamma,\rm iso}$ in the sample defined in \cite{Amati2008} are approximately double for our sample. Using the 1-s peak spectral properties, we also compare the distributions of the $L_{\rm p,iso}$ (Fig.\ref{fig:dis}f) between our sample and the sample defined in \cite{Yonetoku2010}. We find that $L_{\rm p, iso}$ is similar between the samples (see Table \ref{tab:Gaussian}). By using 2-dimensional plots, we also compare the samples in the $E_{\rm p,z}-E_{\gamma, \rm iso}$ and $E_{\rm p,z}-L_{\rm p, iso}$ planes (Figure \ref{fig:comparison}). As shown in Figure \ref{fig:comparison}, the amplitudes and slopes remain similar between our samples and the samples defined in \cite{Amati2008}(Figure \ref{fig:comparison}d) and \cite{Yonetoku2010}(Figure \ref{fig:comparison}e).

In order to perform an evaluation of consistency with previous studies to show quantitative consistency and/or validity between our study and the results present in the \emph{Fermi}/GBM catalog. In Figure \ref{fig:comparison}, we present the parameter comparison between our study and the results in \cite{Poolakkil2021} by using 2-dimensional plots. Three relevant parameters ($E_{\rm p,z}$, $E_{\gamma,\rm iso}$, and $L_{\rm p, iso}$) representing the GRB rest frame properties were selected to be compared. We find that both $E_{\rm p,z}$ (Figure \ref{fig:comparison}a) and $L_{\rm p, iso}$ (Figure \ref{fig:comparison}c) are generally the same. However, $E_{\gamma,\rm iso}$ (Figure \ref{fig:comparison}b) found in \cite{Poolakkil2021} are systemically greater than that in our sample. This may be due to the fact that we have probably selected a slightly different duration from those in \cite{Poolakkil2021}, which is the likely explanation of both $E_{\rm p,z}$ and $L_{\rm p, iso}$ being the same while the $E_{\gamma,\rm iso}$ is different.

\subsection{Possible explanations for the physical origins of the Band-like and CPL-like spectral-energy correlations} 

Despite these spectral-energy correlations have been widely studied in understanding the nature of GRBs, their physical origin is still under debate. Several possible explanations have been proposed in previous studies (i) the result of instrumental selection effects suggested by some authors \citep[e.g.,][]{Nakar2005,Band2005,Butler2007,Kocevski2012}; This possibility was ruled out by some other authors \citep[e.g.,][]{Nava2012} since the time-resolved $E_{\rm p,z}$-$E_{\gamma,\rm iso}$ and $E_{\rm p,z}$-$L_{\gamma, \rm iso}$ correlations found within individual GRBs are similar to each other \citep{Liang2004,Frontera2012,Lu2012,Guiriec2013}, and are also comparable to the $E_{\rm p,z}$-$E_{\gamma,\rm iso}$ and $E_{\rm p,z}$-$L_{\gamma,\rm iso}$ correlation described by the time-integrated spectral properties of different bursts. (ii) Moreover, a recent study \citep{Xue2021} shows in theory the universal correlations among GRBs’ spectral peak, total energy, luminosity and time duration $T_{90}$, provided that GRBs central engines are attributed to gravitational collapses of a massive stellar core or a binary coalescence.

We find that CPL-like GRBs and Band-like GRBs exhibit different Amati and Yonetoku correlations, suggesting that their radiation processes may be different. Several possible scenarios may be used to explain these results. First, due to the different spectral shapes. The spectral shape directly determines $E_{\rm p,z}$ and $E_{\gamma,\rm iso}$ ($L_{\gamma,\rm iso}$), which leads to different spectral-energy correlations. Despite the CPL function being defined as the first part of the Band function, it is still highly debated whether CPL and Band functions have the same physical origins or whether CPL is just an approximation of Band at $\beta \ll 0$ \citep[e.g.,][]{Li2022}. Based on the observed properties of the different distributions of $\beta$ between the Band-like and CPL-like time-resolved spectra using a multipulse GRB sample \citep{Li2021b}, \cite{Li2022} argues that Band and CPL may invoke different radiation mechanisms. Second, individual parameter changes. If CPL is just an approximation of Band at $\beta \ll 0$, this means that both CPL and Band should have the same $E_{\rm p}$ for a given spectrum. This possibility is also supported by some observational evidence (see Figure 9 in \citealt{Li2022}). In this case, $E_{\rm p}$ is equally fixed since it is intrinsically the same in both the models, while $E_{\gamma,\rm iso}$ changes due to the different spectral shapes that occur on the CPL and Band in the high-energy $\beta$ segments, resulting in a different flux integral in the observed energy bands, which we call the $\beta$ effect. For a given observed spectrum, the earlier the spectrum cuts, the less $E_{\rm p}$ will be observed, and the $E^{\rm cut}_{\gamma,\rm iso}$ will be smaller than the $E^{\rm Band}_{\gamma,\rm iso}$ due to $S^{\rm cut}_{\gamma} < S^{\rm Band}_{\gamma}$, where $E^{\rm cut}_{\gamma,\rm iso}$ and $E^{\rm Band}_{\gamma,\rm iso}$ are the $E_{\gamma,\rm iso}$ that integrates the energy flux of the 1-10$^{4}$ keV based on the CPL-like and Band-like spectra, respectively. In contrast, if the cut comes later, the $E_{\rm p}$ would be larger, and $E^{\rm cut}_{\gamma,\rm iso}$ will be closer to $E^{\rm Band}_{\gamma,\rm iso}$. That would result in a flatter increment of the CPL-like bursts than the Band-like bursts as observed in the $E_{\rm p,z}$-$E_{\gamma,\rm iso}$ plane. Alternatively, if the Band-like and CPL-like spectra come from different physical origins, the CPL-$E_{\rm p}$ and Band-$E_{\rm p}$ may not be intrinsically the same, so both $E_{\rm p}$ and $E_{\gamma,\rm iso}$ may be different. In this case, all possible cases ($E_{\rm p,z} \propto E^{\delta}_{\gamma,\rm iso}$ with $\delta_{\rm CPL}<\delta_{\rm Band}$, $\delta_{\rm CPL}>\delta_{\rm Band}$, and $\delta_{\rm CPL}\simeq\delta_{\rm Band}$) in the $E_{\rm p}$-$E_{\gamma,\rm iso}$ plane may be observed. The observations (see Figure \ref{fig:Amati} and Figure \ref{fig:Yonetoku}), however, are only consistent with the first case ($\delta_{\rm CPL}<\delta_{\rm Band}$). Nevertheless, we still cannot rule out the possibility that the CPL and Band originate from different physical origins. In this scenario, a natural interpretation is that the power law of the apparent CPL spectrum is a superposition of the convolution of multiple blackbody spectra in photospheres, and the exponential tail of the apparent CPL spectrum corresponds to the highest temperature of the blackbody spectrum \citep[e.g.,][]{Ryde2010}. The corresponding physical picture is that the photosphere photons observed at a given time interval, corresponding to one time bin in the spectral analysis, are assumed to be emitted from different thin shells, which is given by considering the fireball optical depth falling to a unity. Therefore, the photosphere blackbody spectrum at the given time interval determine the properties of the corresponding shells, and the entire CPL spectrum is conjugated by the photosphere emission from a sequence of such shells.

\section{Conclusions} \label{sec:con}

In this paper, we have studied the model-wise properties of GRB spectral-energy correlations by gathering all {\it Fermi}-detected GRBs with known redshift from July 2008 to May 2022. A complete GRB sample was created, consisting of 153 bursts (17 sGRBs and 136 lGRBs). Our analysis focuses on two important empirical correlations: the relation between the rest-frame peak energy $E_{\rm p,z}$ and isotropic bolometric emission energy $E_{\gamma,\rm iso}$ (the Amati relation), and the relation between the rest-frame peak energy $E_{\rm p,z}$ and peak luminosity $L_{\rm p, iso}$ (the Yonetoku relation). In order to investigate the model-wise properties of GRB spectral-energy correlations, we examined various frequently used spectral models by performing a detailed spectral analysis and model comparisons between various GRB spectral models and their hybrid versions, and all of the $E_{\rm p}$ are obtained from the best-modeled fits. 

Using refined time-integrated spectral analysis and model comparison analysis, we selected 109 GRBs (including 9 sGRBs and 100 lGRBs) with well-measured $E_{\rm p}$ and measured redshift to investigate the model-wise Amati correlation. Via a spectral model-dependent analysis, we found 64 Band-like bursts, 45 CPL-like bursts, 6 Band+BB-like bursts, 5 CPL+BB-like bursts. For the sample as a whole, we found a tight correlation between rest-frame peak energy $E_{\rm p,z}$ and isotropic bolometric emission energy $E_{\gamma,\rm iso}$ ($E_{\rm p,z} \propto E^{0.41\pm0.06}_{\gamma, \rm iso}$) for our Band-like burst sample (sGRBs+lGRBs), while the CPL-like burst sample (sGRBs+lGRBs) in the $E_{\rm p,z}$-$E_{\gamma,\rm iso}$ plane ($E_{\rm p,z} \propto E^{0.00\pm0.13}_{\gamma, \rm iso}$) do not follow the same correlation found in the Band-like burst sample, pointing toward the fact that the Amati correlation is tightly reliant on the model-wise properties. Similar results were also found between the Band-like lGRB sample ($E_{\rm p,z} \propto E^{0.42\pm0.05}_{\gamma, \rm iso}$) and the CPL-like lGRB sample ($E_{\rm p,z} \propto E^{0.25\pm0.12}_{\gamma, \rm iso}$).

By selecting the 1-s peak spectrum, we then repeated our analysis for the Yonetoku correlation as was done for the Amati correlation. We selected 92 GRBs (5 sGRBs and 87 lGRBs) with well-measured $E_{\rm p}$ and peak luminosity $L_{\rm p, iso}$ to investigate the model-wise Yonetoku correlation. Via a spectral model-dependent analysis, we found 55 Band-like bursts, 37 CPL-like bursts, 5 Band+BB-like bursts, 8 CPL+BB-like bursts. We discovered that the Band-like burst samples and the CPL-like burst samples were still well-separated in the $E_{\rm p,z}$-$L_{\gamma,\rm iso}$ plane, which takes the form of $E_{\rm p,z} \propto L^{0.36\pm0.04}_{\rm p, iso}$ for the entire Band-like burst sample (sGRBs+lGRBs) and $E_{\rm p,z} \propto L^{0.40\pm0.12}_{\rm p, iso}$ for the full CPL-like burst sample (sGRBs+lGRBs); and $E_{\rm p,z} \propto L^{0.34\pm0.04}_{\rm p, iso}$ for the Band-like lGRB sample and $E_{\rm p,z} \propto L^{0.40\pm0.08}_{\rm p, iso}$ for the CPL-like lGRB sample, which may not be in line with earlier studies \citep{Yonetoku2010} without distinguishing between the models.

The CPL-like bursts do not fall on the Band-like Amati and Yonetoku relations, suggesting distinct radiation processes. We then discussed several possible explanations, and suggested that the Band-like spectra could originate from a non-thermal emission component (such as synchrotron radiation), that the CPL-like spectra, on the other hand, could be attributed to a superposition of the convolution of multiple blackbody spectra in photospheres, and that the exponential tail of the apparent CPL spectrum corresponds to the highest temperature of the blackbody spectrum. This is also supported by several pieces of additional evidence. For example, an independent $\beta$ distribution is found between the Band-like and CPL-like time-resolved spectral analysis \citep{Li2022}.

We also found three notable outliers (GRB 110721A, GRB 130702A, and GRB 140606B) in the $E_{\rm p,z}$-$E_{\gamma,\rm iso}$ plane and discussed several possibilities, such as whether the selection of the spectral model could have contributed to the outliers in the  $E_{\rm p,z}$-$E_{\gamma,\rm iso}$ plane. In order to maintain the validity of the  $E_{\rm p,z}$-$E_{\gamma,\rm iso}$ correlation, the redshift candidate of $z$ = 3.512 is the more trustworthy between the two redshift candidates ($z$ = 0.382 and $z$ = 3.512) of GRB 110721A as suggested by \cite{Berger2011}. We also found that GRB 130702A is consistent with the existence of a class of nearby and intrinsically faint GRBs with different properties with respect to ``standard" GRBs as first found in \cite{Amati2002}.

\acknowledgments

I thank the anonymous referee for valuable comments and suggestions. I also thank Maria Dainotti, Rahim Moradi, J.-L. Atteia, and the ICRANet members for many helpful discussions on GRB physics and phenomena. This research is made use of the High Energy Astrophysics Science Archive Research Center (HEASARC) Online Service at the NASA/Goddard Space Flight Center (GSFC). 

\vspace{5mm}
\facilities{{\it Fermi}/GBM}
\software{
{\tt 3ML} \citep{Vianello2015}, 
{\tt matplotlib} \citep{Hunter2007}, 
{\tt NumPy} \citep{Harris2020,Walt2011}, 
{\tt SciPy} \citep{Virtanen2020}, 
{\tt $lmfit$} \citep{Newville2016}, 
{\tt astropy} \citep{AstropyCollaboration2013},
{\tt pandas} \citep{Reback2022},
{\tt seaborn} \citep{Waskom2017},
{\tt PyMC3} \citep{Salvatier2016}}  
\bibliography{Myreferences.bib}

\clearpage
\startlongtable
\setlength{\tabcolsep}{-0.1em}
\begin{deluxetable*}{ccccccccc}
\tabletypesize{\scriptsize}
\tablecaption{Global Properties of the Sample}
\tablehead{
\colhead{GRB}
&\colhead{z}
&\colhead{$T_{90}$}
&\colhead{$T_{90}/(1+z)$}
&\colhead{$S_\gamma$}
&\colhead{Detectors}
&\colhead{$[\Delta T_{\rm (bkg,1)},\Delta T_{\rm (bkg,2)}]$}
&\colhead{Averaged Spectrum}
&\colhead{Classified}\\
&&(s)&(s)&(erg.cm$^{-2}$)&&(s)&($E_{\rm p}$)
}
\colnumbers
\startdata
\hline\noalign{\smallskip}
\multicolumn{9}{c}{sGRBs}\tabularnewline
\hline\noalign{\smallskip}
080905(499)&0.1218&0.960$\pm$0.345&0.86$\pm$0.31&(8.50$\pm$0.46)$\times$10$^{-7}$&n3(n6)n7b0&(-20 to -10, 20 to 40)&Unconstrained&Short\\
090510(016)&0.903&0.960$\pm$0.138&0.50$\pm$0.07&(3.37$\pm$0.04)$\times$10$^{-6}$&(n6)n7n9b1&(-20 to -10, 20 to 40)&CPL&Short\\
090927(422)&1.37&0.512$\pm$0.231&0.22$\pm$0.10&(3.03$\pm$0.18)$\times$10$^{-7}$&n2n9(na)b1&(-20 to -10, 40 to 60)&Unconstrained&Short\\
100117A(879)&0.92&0.256$\pm$0.834&0.13$\pm$0.43&(4.23$\pm$0.69)$\times$10$^{-7}$&n3(n4)n8b0&(-20 to -10, 20 to 50)&CPL&Short\\
100206A(563)&0.4068&0.176$\pm$0.072&0.13$\pm$0.05&(7.57$\pm$0.11)$\times$10$^{-7}$&(n0)n1n3b0&(-20 to -10, 30 to 60)&CPL&Short\\
100625A(773)&0.452&0.240$\pm$0.276&0.17$\pm$0.19&(5.63$\pm$0.25)$\times$10$^{-7}$&(n4)b0&(-20 to -10, 20 to 40)&CPL&Short\\
100816A(026)&0.8049&2.045$\pm$0.229&1.13$\pm$0.13&(3.65$\pm$0.05)$\times$10$^{-6}$&n7n8(nb)b1&(-20 to -10, 40 to 60)&Band&Short\\
\enddata 
\vspace{3mm}
\tablecomments{(1) GRB name. (2) Redshift. (3) $t_{90}$, in units of s. (4) The rest-frame $t_{90,z}$, in units of s. (5) The $\gamma$-ray fluence $S_\gamma$, in units of erg cm$^{-2}$. (6) The GBM trigger detectors. (7) The used background intervals. (8) The preferred spectral models for the averaged (time-integrated) spectrum. (9) The classified burst populations. ``Short'': short GRBs; ``S'': single-peak pulse long GRBs; ``M'': multi-peak pulse long GRBs. Note that (i) the background spectrum of GRB 130702A can not be well measured; (ii) In order to obtain an acceptable fit, several special bursts require an additional thermal component either in the time-integrated spectrum (GRB 110721200, GRB 150314205, GRB150403913, and GRB160509374 from the Band+BB model fit and GRB 090902462, GRB 131231198, GRB 181020792, GRB 190114873 from the CPL+BB model fit) or in the 1-s peak spectrum (GRB 090926181, GRB 110721200,GRB 150314205, GRB 150403913, GRB 160625945, GRB 180720598, GRB 210619999 from the Band+BB model fit and GRB 090424592, GRB 090618353, GRB 090902462, GRB 131108862, GRB 170214649, GRB 190114873, GRB 190530430, GRB 201020732, GRB 220527387 from the CPL+BB model), see Table \ref{tab:hybrid} for details.} 
\label{tab:global}
\end{deluxetable*}

\begin{longrotatetable}
\begin{deluxetable}{ccccccccccccccccc}
\setlength{\tabcolsep}{0.0em}
\renewcommand\arraystretch{1.}
\tabletypesize{\tiny}
\tablecaption{Time-integrated Spectral Fit Results \label{tab:Integrated}}
\tablehead{
\specialrule{0em}{1pt}{1pt} 
&\multirow{1}{*}{Time from $T_{0}$}
&\multicolumn{6}{c}{COMP}
&\multicolumn{7}{c}{Band}
&
&\multicolumn{1}{c}{Difference}\\
\cmidrule(lr){2-2} 
\cmidrule(lr){4-9} 
\cmidrule(lr){10-16} 
\cmidrule(lr){17-17}\\
\colhead{ID}
&\colhead{$t_{\rm start} \sim t_{\rm stop}$}
&\colhead{$S$}
&\colhead{$K$}
&\colhead{$\alpha$}
&\colhead{$E_{\rm c}$}
&\colhead{likeli/AIC/BIC}
&\colhead{$E^{\rm rest}_{\rm p}$}
&\colhead{$E_{\gamma,\rm iso}$}
&\colhead{$K$}
&\colhead{$\alpha$}
&\colhead{$E_{\rm p}$}
&\colhead{$\beta$}
&\colhead{likeli/AIC/BIC}
&\colhead{$E^{\rm rest}_{\rm p}$}
&\colhead{$E_{\gamma,\rm iso}$}
&\colhead{$\Delta$AIC}\\
&(s)&&(ph.s$^{-1}$.cm$^{-2}$keV$^{-1}$)&&(keV)&&(keV)&(erg)&(ph.s$^{-1}$.cm$^{-2}$keV$^{-1}$)&&(keV)&&&(keV)&(erg)\\
\hline
}
\colnumbers
\startdata
080916406&-1$\sim$50&28&(0.8$^{+0.2}_{-0.1}$)$\times$10$^{-2}$&-1.04$^{+0.10}_{-0.09}$&124$^{+24}_{-21}$&2729/5464/5475&201$^{+44}_{-39}$&(0.6$^{+0.2}_{-0.1}$)$\times$10$^{52}$&(0.8$^{+0.1}_{-0.1}$)$\times$10$^{-2}$&-1.04$^{+0.09}_{-0.09}$&119$^{+11}_{-12}$&-5.00$^{+0.59}_{-0.08}$&2729/5466/5481&119$^{+11}_{-12}$&(0.6$^{+0.1}_{-0.1}$)$\times$10$^{52}$&2.1\\
081121858&-1$\sim$30&27&(1.4$^{+0.2}_{-0.1}$)$\times$10$^{-2}$&-0.79$^{+0.09}_{-0.09}$&211$^{+40}_{-33}$&2029/4065/4075&899$^{+183}_{-155}$&(6.8$^{+1.8}_{-1.5}$)$\times$10$^{52}$&(1.4$^{+0.1}_{-0.1}$)$\times$10$^{-2}$&-0.79$^{+0.05}_{-0.06}$&255$^{+1}_{-1}$&-4.88$^{+1.21}_{-0.08}$&2029/4067/4081&255$^{+1}_{-1}$&(6.9$^{+0.4}_{-0.4}$)$\times$10$^{52}$&2.0\\
081221681&-1$\sim$40&133&(7.1$^{+0.3}_{-0.3}$)$\times$10$^{-2}$&-0.84$^{+0.03}_{-0.03}$&74$^{+2}_{-2}$&2669/5343/5355&280$^{+13}_{-10}$&(1.2$^{+0.1}_{-0.1}$)$\times$10$^{53}$&(7.3$^{+0.4}_{-0.4}$)$\times$10$^{-2}$&-0.83$^{+0.03}_{-0.03}$&85$^{+2}_{-2}$&-3.81$^{+0.57}_{-0.53}$&2668/5345/5360&85$^{+2}_{-2}$&(1.2$^{+0.1}_{-0.1}$)$\times$10$^{53}$&1.5\\
081222204&-1$\sim$20&57&(2.2$^{+0.2}_{-0.1}$)$\times$10$^{-2}$&-0.90$^{+0.04}_{-0.04}$&147$^{+12}_{-11}$&3059/6124/6137&611$^{+57}_{-53}$&(5.0$^{+0.6}_{-0.5}$)$\times$10$^{52}$&(2.5$^{+0.2}_{-0.2}$)$\times$10$^{-2}$&-0.83$^{+0.05}_{-0.05}$&145$^{+8}_{-8}$&-2.51$^{+0.18}_{-0.18}$&3057/6121/6138&145$^{+8}_{-8}$&(6.0$^{+0.9}_{-0.8}$)$\times$10$^{52}$&-3.2\\
090102122&-1$\sim$40&68&(1.2$^{+0.03}_{-0.03}$)$\times$10$^{-2}$&-1.00$^{+0.02}_{-0.02}$&486$^{+39}_{-34}$&3345/6697/6709&1238$^{+104}_{-92}$&(9.9$^{+0.8}_{-0.7}$)$\times$10$^{52}$&(1.2$^{+0.03}_{-0.03}$)$\times$10$^{-2}$&-1.00$^{+0.02}_{-0.02}$&488$^{+27}_{-26}$&-5.00$^{+0.07}_{-0.01}$&3345/6699/6715&488$^{+27}_{-26}$&(9.9$^{+0.6}_{-0.6}$)$\times$10$^{52}$&2.1\\
\enddata 
\setlength{\parindent}{-9em}
\label{tab:Integrated}
\end{deluxetable}
\end{longrotatetable}

\clearpage
\begin{longrotatetable}
\begin{deluxetable}{ccccccccccccccc}
\setlength{\tabcolsep}{0.10em}
\renewcommand\arraystretch{1.}
\rotate
\tabletypesize{\scriptsize}
\tablecaption{(1-s Peak) Spectral Fit Results \label{tab:1sPeak}}
\tablehead{
\specialrule{0em}{1pt}{1pt} 
&\multirow{1}{*}{Time from $T_{0}$}
&\multicolumn{5}{c}{COMP}
&\multicolumn{6}{c}{Band}
&
&\multicolumn{1}{c}{Difference}\\
\cmidrule(lr){2-2} 
\cmidrule(lr){4-8} 
\cmidrule(lr){9-14} 
\cmidrule(lr){15-15}\\
\colhead{ID}
&\colhead{$t_{\rm start} \sim t_{\rm stop}$}
&\colhead{$S$}
&\colhead{$K$}
&\colhead{$\alpha$}
&\colhead{$E_{\rm c}$}
&\colhead{likeli/AIC/BIC}
&\colhead{$L_{\rm p, iso}$}
&\colhead{$K$}
&\colhead{$\alpha$}
&\colhead{$E_{\rm p}$}
&\colhead{$\beta$}
&\colhead{likeli/AIC/BIC}
&\colhead{$L_{\rm p, iso}$}
&\colhead{$\Delta$AIC}\\
&(s)&&(ph.s$^{-1}$.cm$^{-2}$keV$^{-1}$)&&(keV)&&(erg.s$^{-1}$)&(ph.s$^{-1}$.cm$^{-2}$keV$^{-1}$)&&(keV)&&&(erg.s$^{-1}$)\\
\hline
}
\colnumbers
\startdata
080916406&2$\sim$3&14&(3.7$^{+1.0}_{-0.8}$)$\times$10$^{-2}$&-0.27$^{+0.20}_{-0.21}$&120$^{+30}_{-24}$&701/1408/1420&(0.2$^{+0.1}_{-0.1}$)$\times$10$^{52}$&(3.7$^{+0.9}_{-0.9}$)$\times$10$^{-2}$&-0.27$^{+0.20}_{-0.19}$&207$^{+25}_{-26}$&-5.00$^{+0.28}_{-0.04}$&701/1410/1425&(0.2$^{+0.1}_{-0.1}$)$\times$10$^{52}$&2.1\\
081121858&17$\sim$18&10&(3.2$^{+1.8}_{-1.1}$)$\times$10$^{-2}$&-0.69$^{+0.34}_{-0.32}$&145$^{+88}_{-58}$&507/1019/1030&(4.1$^{+4.7}_{-2.2}$)$\times$10$^{52}$&(3.9$^{+1.8}_{-1.5}$)$\times$10$^{-2}$&-0.58$^{+0.34}_{-0.29}$&165$^{+35}_{-36}$&-2.31$^{+0.37}_{-0.50}$&506/1020/1035&(5.3$^{+4.0}_{-2.8}$)$\times$10$^{52}$&0.6\\
081221681&21$\sim$22&72&(4.5$^{+0.5}_{-0.5}$)$\times$10$^{-1}$&-0.14$^{+0.07}_{-0.07}$&58$^{+3}_{-3}$&749/1505/1516&(1.0$^{+0.2}_{-0.1}$)$\times$10$^{53}$&(5.2$^{+0.7}_{-0.7}$)$\times$10$^{-1}$&-0.06$^{+0.08}_{-0.08}$&103$^{+3}_{-3}$&-3.51$^{+0.29}_{-0.29}$&746/1500/1516&(1.1$^{+0.2}_{-0.2}$)$\times$10$^{53}$&-4.2\\
081222204&3$\sim$4&34&(5.7$^{+0.7}_{-0.6}$)$\times$10$^{-2}$&-0.86$^{+0.08}_{-0.08}$&163$^{+25}_{-22}$&965/1936/1948&(1.0$^{+0.2}_{-0.2}$)$\times$10$^{53}$&(5.7$^{+0.6}_{-0.6}$)$\times$10$^{-2}$&-0.86$^{+0.07}_{-0.07}$&185$^{+15}_{-15}$&-5.00$^{+0.01}_{-0.01}$&965/1938/1954&(1.0$^{+0.2}_{-0.2}$)$\times$10$^{53}$&2.1\\
090102122&14$\sim$15&34&(5.5$^{+0.3}_{-0.3}$)$\times$10$^{-2}$&-0.16$^{+0.08}_{-0.08}$&214$^{+21}_{-18}$&891/1788/1801&(5.6$^{+1.2}_{-1.1}$)$\times$10$^{52}$&(5.5$^{+0.3}_{-0.4}$)$\times$10$^{-2}$&-0.15$^{+0.08}_{-0.09}$&394$^{+23}_{-21}$&-4.52$^{+2.05}_{-0.04}$&891/1790/1807&(6.1$^{+1.6}_{-0.8}$)$\times$10$^{52}$&2.0\\
090323002&66$\sim$67&21&(6.2$^{+0.7}_{-0.6}$)$\times$10$^{-2}$&-0.27$^{+0.11}_{-0.11}$&193$^{+29}_{-26}$&1415/2837/2849&(3.5$^{+1.3}_{-0.9}$)$\times$10$^{53}$&(6.3$^{+0.6}_{-0.6}$)$\times$10$^{-2}$&-0.26$^{+0.11}_{-0.11}$&333$^{+28}_{-30}$&-4.13$^{+1.88}_{-0.35}$&1415/2839/2855&(3.9$^{+1.2}_{-0.7}$)$\times$10$^{53}$&2.0\\
\enddata 
\vspace{3mm}
\setlength{\parindent}{-9em}
\tablecomments{Same as Table \ref{tab:Integrated} but for the 1-s peak spectrum.}\label{tab:1speak}
\end{deluxetable}
\end{longrotatetable}

\clearpage
\setlength{\tabcolsep}{0.05em}
\begin{deluxetable}{cccccccccccccc}
\centering
\rotate
\tabletypesize{\tiny}
\tablecaption{Time-integrated Spectral Fit Results for those GRBs required an Additional Model \label{tab:BB}}
\tablehead{
\specialrule{0em}{3pt}{3pt} 
& 
&
&
&\multicolumn{4}{c}{Standard Model}
&\multicolumn{2}{c}{Additional Model}\\
\cmidrule(lr){5-8} \cmidrule(lr){9-10}\\
&\multirow{1}{*}{Time from $T_{0}$}&&\multirow{1}{*}{Model}&\multicolumn{4}{c}{Band or COMP}&\multicolumn{2}{c}{Blackbody}&\colhead{Thermal Flux}&\colhead{Total Flux}&\colhead{Information criteria}&\multicolumn{1}{c}{Difference}\\
\cmidrule(lr){2-2} \cmidrule(lr){4-4} \cmidrule(lr){5-8} \cmidrule(lr){9-10} \cmidrule(lr){11-11} \cmidrule(lr){12-12} \cmidrule(lr){13-13} \cmidrule(lr){14-14}\\
\colhead{ID}
&\colhead{$T_{\rm start} \sim T_{\rm stop}$}
&\colhead{$S$}
&\colhead{Parameter}
&\colhead{$K$}
&\colhead{$\alpha$}
&\colhead{$E_{\rm p}$ or $E_{\rm c}$}
&\colhead{$\beta$}
&\colhead{$K$}
&\colhead{k$T$}
&\colhead{$F_{\rm BB}$}
&\colhead{$F_{\rm tot}$}
&\colhead{Likeli/AIC/BIC}
&\colhead{$\Delta$BIC}\\
}
\colnumbers
\startdata
\hline\noalign{\smallskip}
\multicolumn{14}{c}{Time-intergrated spectral analysis}\tabularnewline
\hline\noalign{\smallskip}
090618353&-1$\sim$110&196&CPL+BB&(3.1$^{+0.1}_{-0.1}$)$\times$10$^{-2}$&-1.45$^{+0.02}_{-0.02}$&729$^{+77}_{-72}$&\nodata&(1.5$^{+0.1}_{-0.1}$)$\times$10$^{-4}$&21$^{+1}_{-1}$&(2.9$^{+0.4}_{-0.4}$)$\times$10$^{-7}$&(2.6$^{+0.2}_{-0.2}$)$\times$10$^{-6}$&2501/5013/5030&-348.6\\
090902462&-1$\sim$20&271&CPL+BB&(8.1$^{+0.1}_{-0.1}$)$\times$10$^{-2}$&-1.15$^{+0.01}_{-0.01}$&1228$^{+54}_{-55}$&\nodata&(0.7$^{+0.1}_{-0.1}$)$\times$10$^{-6}$&158$^{+3}_{-3}$&(4.8$^{+0.5}_{-0.5}$)$\times$10$^{-6}$&(1.6$^{+0.1}_{-0.1}$)$\times$10$^{-5}$&5068/10147/10168&-925.0\\
110721200&-1$\sim$20&92&Band+BB&(1.8$^{+0.1}_{-0.1}$)$\times$10$^{-2}$&-1.20$^{+0.03}_{-0.03}$&1392$^{+313}_{-302}$&-2.10$^{+0.11}_{-0.11}$&(9.6$^{+2.4}_{-2.4}$)$\times$10$^{-6}$&33$^{+3}_{-3}$&(1.2$^{+0.6}_{-0.4}$)$\times$10$^{-7}$&(3.9$^{+0.6}_{-0.5}$)$\times$10$^{-6}$&3117/6245/6270&-7.9\\
131231198&0.06$\sim$40&355&CPL+BB&(5.7$^{+0.1}_{-0.1}$)$\times$10$^{-2}$&-1.40$^{+0.01}_{-0.01}$&566$^{+32}_{-30}$&\nodata&(1.2$^{+0.1}_{-0.1}$)$\times$10$^{-4}$&23$^{+0}_{-0}$&(3.1$^{+0.4}_{-0.3}$)$\times$10$^{-7}$&(4.1$^{+0.2}_{-0.2}$)$\times$10$^{-6}$&4558/9125/9146&-331.2\\
150314205&-1$\sim$15&210&Band+BB&(9.2$^{+0.4}_{-0.3}$)$\times$10$^{-2}$&-0.72$^{+0.02}_{-0.02}$&414$^{+15}_{-16}$&-3.20$^{+0.35}_{-0.34}$&(2.8$^{+0.8}_{-0.8}$)$\times$10$^{-5}$&29$^{+3}_{-3}$&(2.0$^{+1.1}_{-0.8}$)$\times$10$^{-7}$&(6.7$^{+0.6}_{-0.5}$)$\times$10$^{-6}$&2344/4701/4724&-14.5\\
150403913&-1$\sim$30&118&Band+BB&(2.1$^{+0.1}_{-0.1}$)$\times$10$^{-2}$&-1.04$^{+0.02}_{-0.02}$&851$^{+62}_{-61}$&-2.74$^{+0.22}_{-0.21}$&(2.8$^{+0.5}_{-0.4}$)$\times$10$^{-5}$&26$^{+1}_{-1}$&(1.3$^{+0.3}_{-0.3}$)$\times$10$^{-7}$&(2.9$^{+0.2}_{-0.2}$)$\times$10$^{-6}$&2592/5196/5219&-115.6\\
\enddata 
\vspace{3mm}
\tablecomments{(1) GRB name. (2) The selected source interval $T_{\rm start} \sim T_{\rm stop}$ (in units of s). (3) The corresponding significance $S$. (4) The used hybrid model. (5)-(8) The best-fit parameters for the Band or COMP model: normalization $K$ (in units of ph.s$^{-1}$.cm$^{-2}$keV$^{-1}$), low-energy power-law index $\alpha$, peak energy $E_{\rm p}$ (or cut-off energy $E_{\rm c}$) of the $\nu F_{\nu}$ spectrum (in units of keV), and high-energy power-law index $\beta$. (9)-(10) The best-fit parameters for the additional model: normalization $K$ (in units of ph.s$^{-1}$.cm$^{-2}$keV$^{-1}$), and blackbody temperature k$T$ (in units of keV); the thermal ($\nu F_{\nu}$) energy flux $F_{\rm BB}$ (in units of erg cm$^{-2}$ s$^{-1}$) and the total ($\nu F_{\nu}$) energy flux $F_{\rm tot}$ (in units of erg cm$^{-2}$ s$^{-1}$); the likelihood and the statistical information criteria (AIC and BIC); and the difference of BIC between Band+BB (or CPL+BB) and Band-alone (or CPL-alone): BIC=$BIC_{\rm Band+BB}$-$BIC_{\rm Band}$ or $BIC_{\rm CPL+BB}$-$BIC_{\rm CPL}$.}
\label{tab:hybrid}
\end{deluxetable}

\clearpage
\begin{table*}
\refstepcounter{table}\label{tab:Gaussian}
\setlength{\tabcolsep}{0.10em}
\renewcommand\arraystretch{1.2}
{\bf Table~5~~ Results of the Average and Deviation Values of the Parameter Distribution}\\
\centering
\scalebox{0.80}{
\begin{tabular}{c|ccc}
\hline
&Sample size & Gaussian fit result&Source\\
\hline
{\bf Global properties}\\
\hline
Observed-frame duration [$T_{90}$]&153&$\rm log_{10}$(41.20)$\pm$0.49~s&this paper\\
Rest-frame duration [$T_{90,z}$]&153&$\rm log_{10}$(15.16)$\pm$0.57~s&this paper\\
$k_{\rm c}$&103&0.95$\pm$0.11&this paper\\
Bolometric (1-10$^{4}$ keV) fluence [$S^{\rm Bolometric}_{\rm \gamma}$]&103&$\rm log_{10}$(2.99$\times$10$^{5}$)$\pm$0.67 erg cm$^{-2}$&this paper\\
GBM (8 keV-40 MeV) fluence [$S^{\rm GBM}_{\rm \gamma}$]&103&$\rm log_{10}$(3.01$\times$10$^{5}$)$\pm$0.67 erg cm$^{-2}$&this paper\\
Time-integrated spectra, peak energy [$E_{\rm p,z}$]&117&$\rm log_{10}$(533)$\pm$0.40 keV&this paper\\
\hline
\end{tabular}
}
\end{table*}

\clearpage
\begin{deluxetable*}{ccccccc}
\tablewidth{0pt}
\tablecaption{Results of Linear Regression Analysis for the Spectral-energy Relations}\label{tab:Spearman}
\tablehead{
\colhead{Sample}
&\colhead{Preferred Spectral model}
&\colhead{Number}
&\colhead{Expression}
&\colhead{$E^{\rm piv}_{\gamma,\rm iso}$ or $L^{\rm piv}_{\rm p, iso}$}
&\colhead{$R$}
&\colhead{$p$}\\
}
\startdata
\hline\noalign{\smallskip}
\multicolumn{6}{c}{$E_{\rm p,z}$-$E_{\gamma,\rm iso}$ (Amati) Correlation}\tabularnewline
\hline\noalign{\smallskip}
Band-like lGRBs&Band&61&${\it E_{\rm p,z} {\rm /(keV)}}=(229\pm29)\left(\frac{E_{\gamma,\rm iso}{\rm /(erg)}}{E^{\rm piv}_{\gamma,\rm iso}}\right)^{(0.42\pm0.05)}$&2$\times$10$^{52}$erg&0.71&$<$10$^{-4}$\\
CPL-like lGRBs&CPL&39&${\it E_{\rm p,z} {\rm /(keV)}}=(341\pm65)\left(\frac{E_{\gamma,\rm iso}{\rm /(erg)}}{E^{\rm piv}_{\gamma,\rm iso}}\right)^{(0.25\pm0.12)}$&2$\times$10$^{52}$erg&0.11&0.67\\
All Band-like bursts (sGRBs+lGRBs)&Band&64&${\it E_{\rm p,z} {\rm /(keV)}}=(235\pm36)\left(\frac{E_{\gamma,\rm iso}{\rm /(erg)}}{E^{\rm piv}_{\gamma,\rm iso}}\right)^{(0.41\pm0.06)}$&2$\times$10$^{52}$erg&-0.07&0.74\\
All CPL-like bursts (sGRBs+lGRBs)&CPL&45&${\it E_{\rm p,z} {\rm /(keV)}}=(336\pm83)\left(\frac{E_{\gamma,\rm iso}{\rm /(erg)}}{E^{\rm piv}_{\gamma,\rm iso}}\right)^{(0.00\pm0.13)}$&2$\times$10$^{52}$erg&0.86&$<$10$^{-4}$\\
\hline\noalign{\smallskip}
\multicolumn{6}{c}{$E_{\rm p,z}$-$L_{\rm p,iso}$ (Yonetoku) Correlation}\tabularnewline
\hline\noalign{\smallskip}
Band-like lGRBs&Band&53&${\it E_{\rm p,z} {\rm /(keV)}}=(307\pm34)\left(\frac{L_{\rm p, iso}{\rm /(erg.s^{-1})}}{L^{\rm piv}_{\rm p, iso}}\right)^{(0.34\pm0.04)}$&2$\times$10$^{52}$ erg s$^{-1}$&0.83&$<$10$^{-4}$\\
CPL-like lGRBs&CPL&35&${\it E_{\rm p,z} {\rm /(keV)}}=(454\pm55)\left(\frac{L_{\rm p, iso}{\rm /(erg.s^{-1})}}{L^{\rm piv}_{\rm p, iso}}\right)^{(0.40\pm0.08)}$&2$\times$10$^{52}$ erg s$^{-1}$&0.69&$<$10$^{-4}$\\
\enddata
\label{tab:Gaussian}
\end{deluxetable*}

\clearpage

\clearpage

\clearpage
\begin{figure*}[ht!]
\includegraphics[width=1\textwidth]{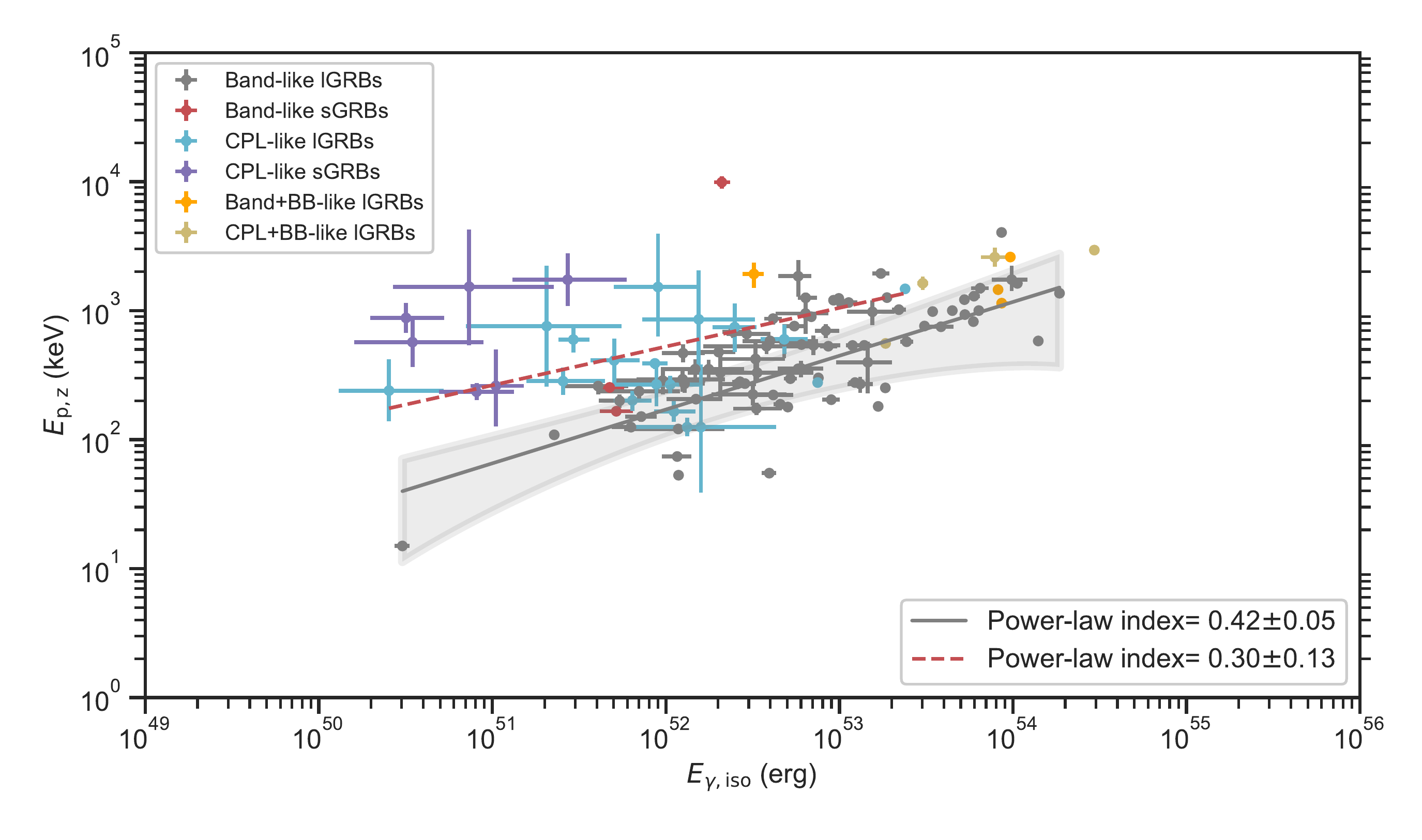}
\caption{109 GRBs with firm estimates of redshift and well-measured $E_{\rm p}$, plotted in the $E_{\rm p,z}$-$E_{\gamma,\rm iso}$ plane. Data points with magenta, cyan, and orange indicate the Band-like bursts, CPL-like bursts, Band+BB-like bursts, and CPL+BB-like bursts, respectively. The solid line (grey) is the best fit using the power-law model with $2\sigma$ (95\% confidence interval) error region (shadow area) for the Band-like lGRBs while the red dashed line represents the best fit for the CPL-like lGRBs.}
\label{fig:Amati}
\end{figure*}

\clearpage
\begin{figure*}[ht!]
\includegraphics[width=1\textwidth]{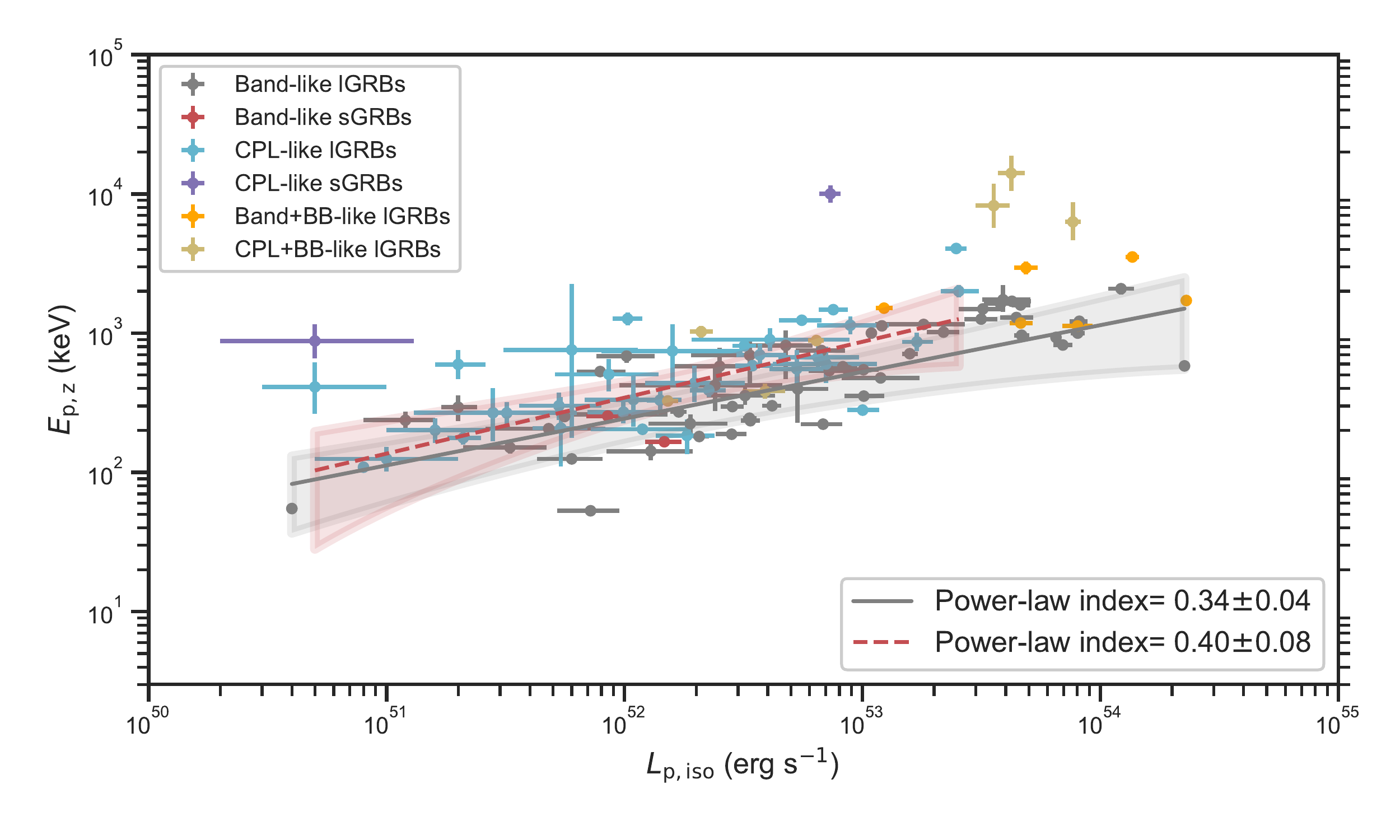}
\caption{Same as Figure \ref{fig:Amati} but for the Yonetoku ($E_{\rm p,z}$-$L_{\rm p, iso}$) correlation analysis using 92 GRBs with firm estimates of redshift and well-measured $E_{\rm p}$ from the time-integrated spectral analysis and the peak luminosity $L_{\rm p, iso}$ from the 1-s peak spectral analysis. The symbols and colors are the same as Figure \ref{fig:Amati}.}
\label{fig:Yonetoku}
\end{figure*}

\clearpage
\begin{figure}[ht!]
\includegraphics[width=1\textwidth]{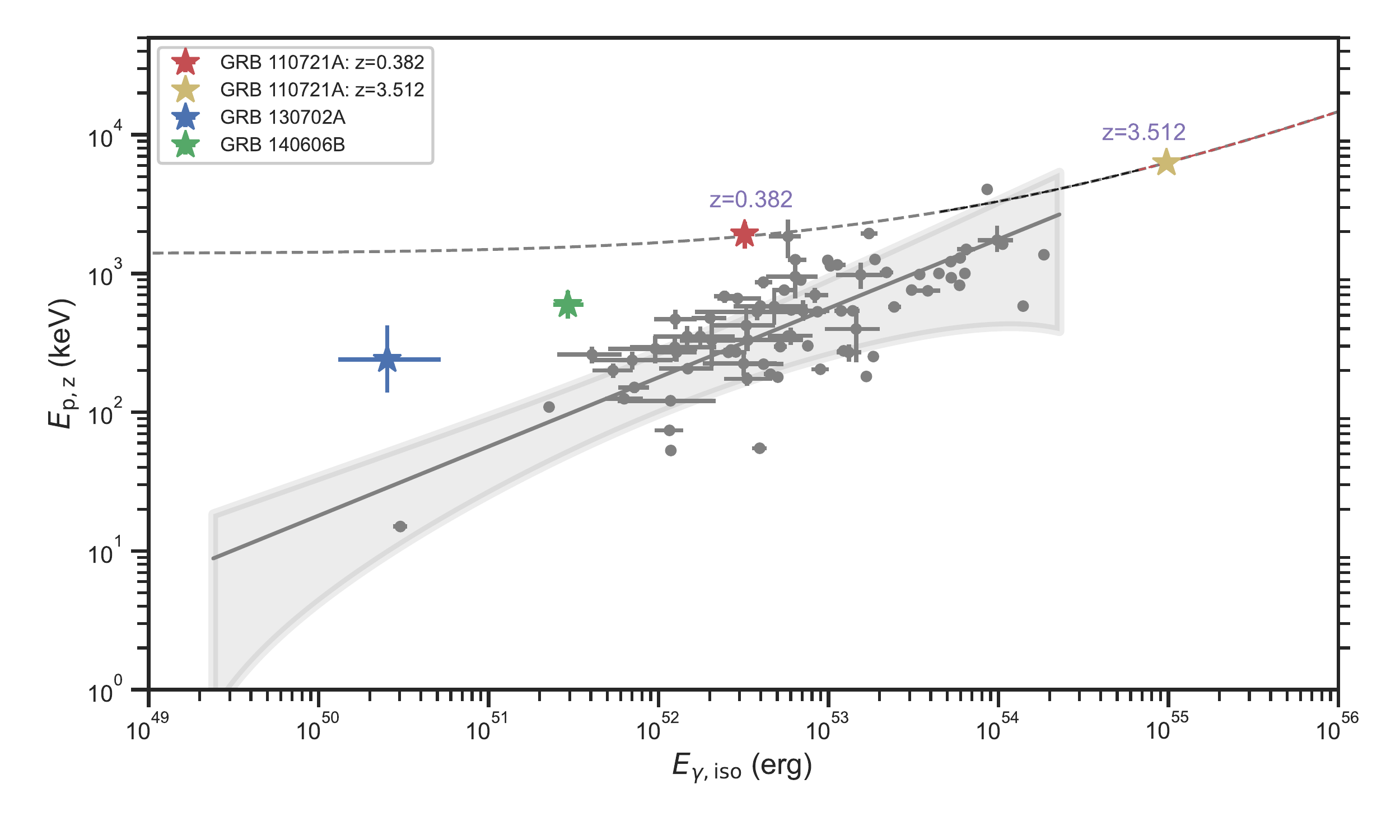}
\caption{Three outliers (star data points) and 64 lGRBs (circular data points) in the $E_{\rm p,z}$-$E_{\gamma,\rm iso}$ plane. The dash lines represent the redshift evolution in the $E_{\rm p,z}$-$E_{\gamma,\rm iso}$ plane for the three outliers. Colors with grey black and red denote the low (from 0.01 to 1), intermediate (from 1 to 3), and high redshift (from 3 to 10) regions.}
\label{fig:outlier}
\end{figure}

\clearpage
\begin{figure}[ht!]
\includegraphics[width=1\textwidth]{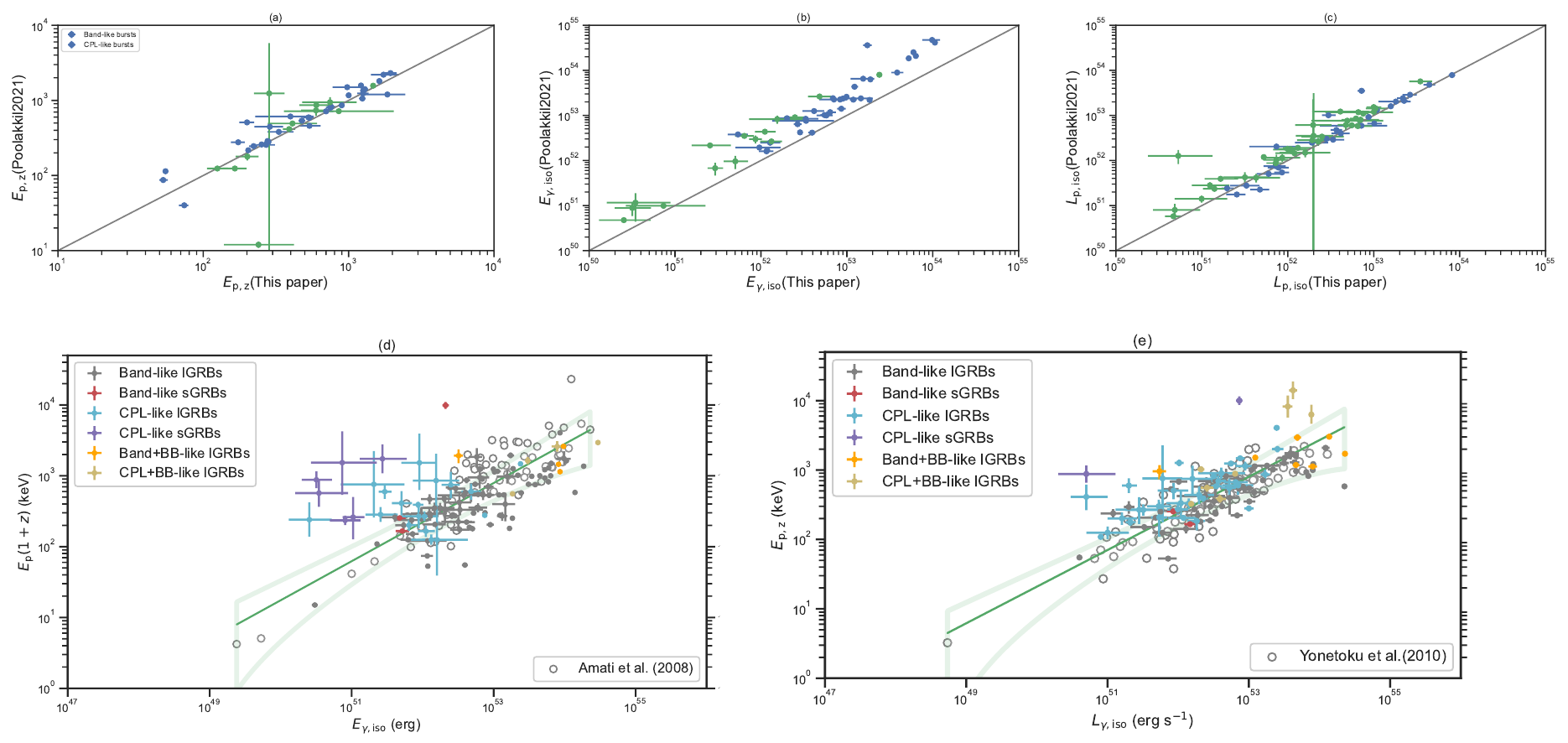}
\caption{{\bf Upper panel}: Parameter comparison ($E_{\rm p,z}$, $E_{\gamma,\rm iso}$, and $L_{\rm p, iso}$) between our study and \cite{Poolakkil2021} by using 2-dimensional plots. Different color represents different groups: blue color (Band-like bursts) and green color (CPL-like bursts). {\bf Lower panel}: Different burst samples from the $E_{\rm p,z}$-$E_{\gamma,\rm iso}$ and $E_{\rm p,z}$-$L_{\rm p, iso}$ planes are compared. The data points indicated by hollow points represent the sample defined in \cite{Amati2008} and \cite{Yonetoku2010}. The solid green lines are the best fit using the power-law model with $2\sigma$ (95\% confidence interval) error region (shadow area) to the sample defined in \cite{Amati2008} and \cite{Yonetoku2010}, respectively. The data points indicated by the other colors (the same as Figure \ref{fig:Amati}) represent the burst samples defined in this paper based on their model-wise properties.}
\label{fig:comparison}
\end{figure}

\clearpage
\appendix
\setcounter{figure}{0}    
\setcounter{section}{0}
\setcounter{table}{0}
\renewcommand{\thesection}{A\arabic{section}}
\renewcommand{\thefigure}{A\arabic{figure}}
\renewcommand{\thetable}{A\arabic{table}}
\renewcommand{\theequation}{A\arabic{equation}}

In this appendix, we provide additional figures and tables.

\section{Correlation coefficient by Markov chain Monte Carlo algorithm}

\begin{figure*}[ht!]
\includegraphics[width=1\textwidth]{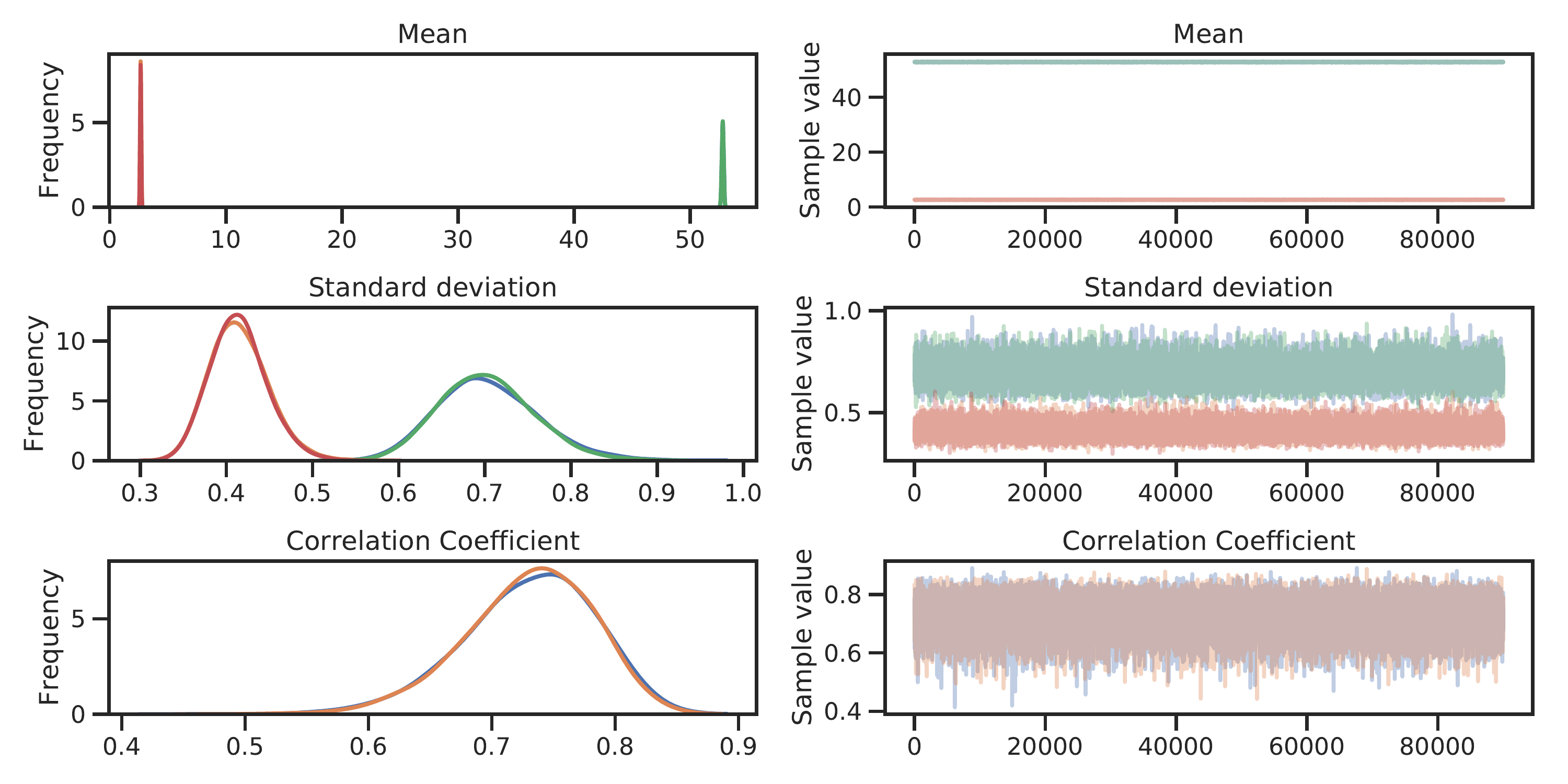}
\caption{MCMC iteration for the mean values and standard deviation of the $E_{\rm p}$ and $E_{\gamma,\rm iso}$, and their correlation coefficient for our Band-like lGRB sample, including the value of 10$^{5}$ iterations (right) and their distribution (left).}
\label{fig:mcmc}
\end{figure*}

The correlation coefficient is used to assess the reliability of the correlation between the cosmological rest-frame peak energy ($E_{\rm p}$) and either the isotropic-bolometric-equivalent emission energy ($E_{\gamma,\rm iso}$) or the isotropic-bolometric-equivalent peak luminosity ($L_{\rm p, iso}$) for our samples. By implementing the Python package {\tt PyMC3} \citep{Salvatier2016}, we use a normal-LKJ correlation prior distribution and the Markov chain Monte Carlo (MCMC) algorithm to obtain the covariance matrix of the multivariate normal distribution, and the correlation coefficient between the parameters is obtained as a result \citep{chib2001markov} by iterating 10$^{5}$ times and burning the first 10$^{4}$ times of the MCMC samples. Our results are summarized in Table \ref{tab:MCMC}, including the expected values ($\mu_1$) and standard deviation ($\sigma_1$) of the normal distribution of the analyzed parameters, and their correlation coefficients, as well as the related Highest Density Interval (HDI) of the posterior distributions, ranging from 3\% to 97\%. In Figure \ref{fig:mcmc}, using the Band-like lGRB sample as an example, we show an MCMC iteration for the mean values and standard deviation of the $E_{\rm p}$ and $E_{\gamma,\rm iso}$, and their correlation coefficient, including the value of 10$^{5}$ iterations (right) and their distribution (left).

\clearpage
\begin{deluxetable*}{ccccccc}
\label{tab:MCMC}
\tablewidth{0pt}
\setlength\tabcolsep{5pt} 
\tabletypesize{\scriptsize}
\tablecaption{Results of correlation coefficient of Markov chain Monte Carlo algorithm.}
\tablehead{
\colhead{Sample}
&\colhead{$\mu_1$}
&\colhead{$\sigma_1$}
&\colhead{$\mu_2$}
&\colhead{$\sigma_2$}
&\colhead{Correlation Coefficient}
&\colhead{hdi interval}\\
&\multicolumn{2}{c}{$\rm log_{10}$($E_{\gamma,\rm iso}$ or $L_{\rm p, iso}$)}
&\multicolumn{2}{c}{$\rm log_{10}$($E_{\rm p,z}$)}
&\colhead{($E_{\gamma,\rm iso}$-$E_{\rm p,z}$) or ($L_{\rm p, iso}$-$E_{\rm p,z}$)}
&\colhead{[3\% to 97\%]}\\
\cmidrule(lr){2-3} \cmidrule(lr){4-5}
}
\startdata
\hline\noalign{\smallskip}
\multicolumn{5}{c}{$E_{\rm p,z}$-$E_{\gamma,\rm iso}$ (Amati) Correlation}\tabularnewline
\hline\noalign{\smallskip}
Band-like lGRBs&52.80$\pm$0.08&0.70$\pm$0.06&2.65$\pm$0.05&0.41$\pm$0.03&0.73$\pm$0.05&[0.62 to 0.83]\\
CPL-like lGRBs &51.99$\pm$0.16&0.70$\pm$0.11&2.59$\pm$0.08&0.37$\pm$0.07&0.22$\pm$0.22&[-0.22 to 0.63]\\
All Band-like bursts (sGRBs+lGRBs)&52.78$\pm$0.08&0.73$\pm$0.06&2.67$\pm$0.05&0.44$\pm$0.03&0.68$\pm$0.06&[0.55 to 0.78]\\
All CPL-like bursts (sGRBs+lGRBs)&51.72$\pm$0.16&0.79$\pm$0.10&2.65$\pm$0.08&0.38$\pm$0.06&-0.01$\pm$0.20&[-0.40 to 0.37]\\
\hline\noalign{\smallskip}
\multicolumn{5}{c}{$E_{\rm p,z}$-$L_{\rm p,iso}$ (Yonetoku) Correlation}\tabularnewline
\hline\noalign{\smallskip}
Band-like lGRBs&52.66$\pm$0.13&0.79$\pm$0.09&2.77$\pm$0.07&0.45$\pm$0.05&0.64$\pm$0.10&[0.43 to 0.82]\\
CPL-like lGRBs&52.15$\pm$0.12&0.83$\pm$0.08&2.70$\pm$0.05&0.31$\pm$0.03&0.70$\pm$0.08&[0.54 to 0.83]\\
All Band-like bursts (sGRBs+lGRBs)&52.64$\pm$0.12&0.77$\pm$0.09&2.78$\pm$0.08&0.48$\pm$0.06&0.61$\pm$0.10&[0.41 to 0.80]\\
All CPL-like bursts (sGRBs+lGRBs)&52.09$\pm$0.12&0.85$\pm$0.07&2.70$\pm$0.04&0.30$\pm$0.03&0.65$\pm$0.08&[0.49 to 0.81]\\
\enddata
\end{deluxetable*}

\end{document}